# Brain Tumor Detection Based on a Novel and High-Quality Prediction of the Tumor Pixel Distributions

**Yanming Sun and Chunyan Wang**

Department of Electrical and Computer Engineering, Concordia University,

1455 De Maisonneuve Blvd. W. Montreal, Quebec, Canada, H3G 1M8

Corresponding author: Chunyan Wang (chunyan@ece.concordia.ca)

## Abstract

In this paper, we propose a system to detect brain tumor in 3D MRI brain scans of Flair modality. It performs 2 functions: (a) predicting gray-level and locational distributions of the pixels in the tumor regions and (b) generating tumor mask in pixel-wise precision. To facilitate 3D data analysis and processing, we introduced a 2D histogram presentation that comprehends the gray-level distribution and pixel-location distribution of a 3D object. In the proposed system, particular 2D histograms, in which tumor-related feature data get concentrated, are established by exploiting the left-right asymmetry of a brain structure. A modulation function is generated from the input data of each patient case and applied to the 2D histograms to attenuate the element irrelevant to the tumor regions. The prediction of the tumor pixel distribution is done in 3 steps, on the axial, coronal and sagittal slice series, respectively. In each step, the prediction result helps to identify/remove tumor-free slices, increasing the tumor information density in the remaining data to be applied to the next step. After the 3-step removal, the 3D input is reduced to a minimum bounding box of the tumor region. It is used to finalize the prediction and then transformed into a 3D tumor mask, by means of gray level thresholding and low-pass-based morphological operations. The final prediction result is used to determine the critical threshold. The proposed system has been tested extensively with the data of more than one thousand patient cases in the datasets of BraTS 2018~21. The test results demonstrate that the predicted 2D histograms have a high degree of similarity with the true ones. The system delivers also very good tumor detection results, comparable to those of state-of-the-art CNN systems with mono-modality inputs, which is achieved at an extremely low computation cost and no need for training.

**Keywords:** brain tumor detection, image processing, 3D MRI brain image processing, deterministic model, prediction of object-pixel distribution, tumor mask generation

# 1. Introduction

Brain tumor detection is important for brain cancer diagnosis. Manual detection is a time-consuming task performed by medical specialists, and the limitation in resources may delay the detection and diagnosis. Developing compute vision systems for fully automatic brain tumor detection can help to reduce the work load of the medical specialists and improve the chance of timely diagnosis and treatments.

A brain image is often generated by magnetic resonance imaging (MRI) and its 3D structure is presented as a series of 2D slices. Two examples of image slices, produced by MRI of Flair modality, are illustrated in Fig. 1 (a) and (c), together with their tumor mask slices shown in Fig. 1 (b) and (d). A brain tumor detection system is expected to generate a 3D tumor mask covering, in a pixel-wise precision, the entire tumor region in all the slices of the series.

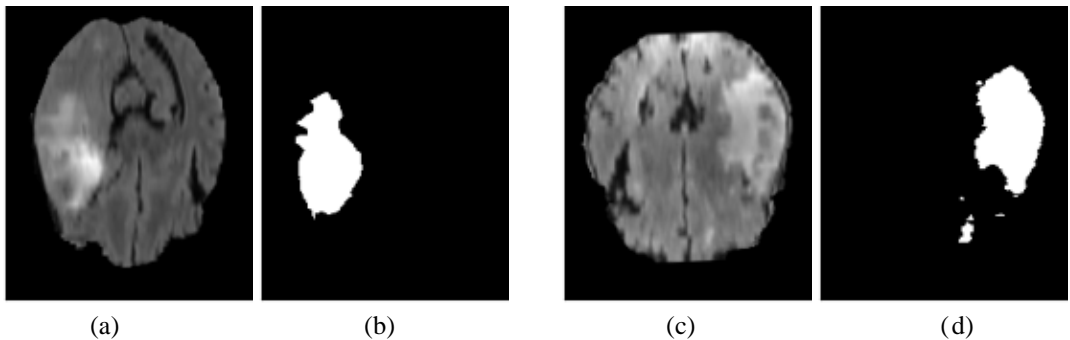

    (a)        (b)        (c)        (d)

Fig. 1 (a)(c) Slices sampled from two 3D brain images of Flair modality.
    (b)(d) Binary tumor masks of the slices in (a) and (c).

Tumors can be found everywhere in a brain and appear in very different sizes, shapes and texture patterns. They can change from case to case with little coherence, making the detection by computer vision a very challenging task. Such a task seems to be too complex to be handled by conventional filtering systems.

Convolutional neural network (CNN) can be a solution to the brain tumor detection problem, as it can have many filtering layers and a very large number of filters in each layer to deal with enormous variations in its objects. Various CNN structures have been developed for particular kinds of tasks. For example, Visual Geometry Group (VGG) [1] is used to extract various image features, and many CNNs for medical image segmentation/detection are built on the basis of U-Net structure [2][3][4][5][6][7]. The processing power of a CNN is related to its number of parameters in the filtering kernels, and the values of these parameters are determined by means of a training process. It should, however, be noted that the quantity and quality of data samples should match the number of trainable parameters to train the CNN decently. If one attempts to have more processing power by increasing the number of kernels/parameters of a CNN, more data samples will be needed and the data resources may not necessarily meet the growing needs. Hence, the problem of insufficient data samples is yet to be overcome one way or another.

It should also be noted that operating a CNN requires a large amount of computation, let alone training it. Though the computation capacity is constantly getting improved in recently years, the pace of growing may not match what is desired. Moreover, the accessibility of the existing computation resources is another issue. Hence, maximizing the computation efficiency, i.e., performing the same task while using the minimum amount of computation, should still be a point drawing the attention of designers.

The objective of the work presented in this paper is to develop a computation-efficient system to detect brain tumors in 3D mono-modality MRI scans. It is non-CNN, able to find tumor information from its input data, and no training is needed.

To achieve a high processing quality at the lowest computation cost, the system has a very specific structure. The main processing in the system is to predict, step-by-step, the gray level distribution of the pixels in the tumor regions of a 3D brain image. The prediction result of each step is used to detect and to remove regions of non-interest, i.e., tumor-free regions, from the 3D image. Each removal reduces the data volume and improves the density of the tumor information, facilitating the prediction in the succeeding step. The final prediction result is then applied to the remaining 3D data to detect, by means of very simple operations, the tumor locations precisely. In the design of this system, the following ideas and methods have been proposed and implemented, as positive contributions to this topic area.

- Presentation of 2D histogram of 3D data. It encompasses the gray level distribution of the data and their locational distribution. As the pixels of a 3D image can be presented in a series of 2D slices, the 2D histogram illustrates how the pixels at a particular gray level, or in a given gray level range, are distributed over the slices.

- Histogram modulation function to attenuate the presence of tumor-free elements. It transforms a histogram representing the gray level distribution of the elements in both tumoral and tumor-free areas to a histogram representing mainly the distribution of the tumoral elements. The modulation function is generated with the original data of each patient case so that its characteristic can adapt to the data distribution of the particular case.

- Method to interleave a step-by-step processing to predict the gray level distribution of pixels in the object region and that to detect/remove non-object region in the same 3D image. The two interact with each other and complement to each other: The result of each prediction step is used to detect and to remove non-object regions, improving the density of the object information and benefiting the prediction in the following step.

This paper consists of 6 sections. The related work is presented in Section 2. The proposed 2D histogram method is described in Section 3. The design of the proposed system is presented in Section 4, and the 2D histogram method is applied throughout the design process. Section 5 is dedicated to the presentation of the experiment results. A conclusion is presented in Section 6.

## 2. Related Work

A brain tumor detection involves, usually, feature extraction from the input data and classification operation applied to the extracted features. As the 2 kinds of operations can be performed in different ways, there are varieties systems reported in this topic area and relevant to the work presented in this paper.

Feature extraction can be performed by means of filters. Gabor filters are commonly used in non-CNN systems where filtering kernels are not determined by training. The extracted features are then applied to various classifiers. For example, the method of Extremely Randomized Trees can be used for this purpose [8][9]. In some systems, Gabor filtering method is combined with Support Vector Machine (SVM) to detect brain tumors [10][11]. One can also combine Gabor filtering and K-means clustering methods for feature extraction and SVM together with Random Forest (RF) for classification to improve the detection result [12]. Gabor filtering and Walsh-Hadamard transform (WHT) can be used for feature extraction, and Fuzzy C-Means clustering for classification [13].

Some region-based image segmentation methods are used to detect brain tumor, e.g., homogeneity- and object-feature based Random Walks (HORW) [14], visual saliency & automated grow-cut [15] and multi-agent adaptive region grow [16]. In these methods, initial seed points should be selected, and the neighboring pixels are examined and determined whether they belong to the same region of the seed.

The feature information concerning brain tumors can also be extracted by measuring asymmetry of a brain structure as a tumor can make its left-right halves less symmetrical. The degree of asymmetry of the 2 halves can be represented by their dissimilarity, measured by calculating, for example, the pixel-by-pixel difference of the two 3D halves [3][17] or the correlation between them [18]. The 3D data resulting from such a calculation are used as feature data to be applied to a classifier of Random Forest [17]. In the symmetric driven generative adversarial network (SD-GAN) [3], the left and right halves of a tumor-free brain image are reconstructed and the brain tumor segmentation is based on high reconstruction errors arising from asymmetry. In the deep convolutional symmetric neural network (DCSNN) [18], the 3D data representing the correlation between the left and right halves are used to generate symmetric masks applied to the feature maps of convolution layers to enhance the tumor-related features.

It should be noted that the 3D data produced by the dissimilarity measures represent all the asymmetry, caused not only by the tumors but also by the differences of texture details in healthy parts. To make the latter less pronounced, one can measure the degree of asymmetry of the 2 halves based on their statistical presentations, e.g., gray level distributions, instead of their 3D data. For example, the difference between the gray level histograms of the normal hemisphere and the pathological hemisphere of a brain are calculated by a very simple subtraction operation [19]. One can use the pixels in a 3D image to generate multiple pairs of histograms, each of which given by 2 subregions located symmetrically in the 2 halves, and calculate the degree of dissimilarity by Bhattacharya coefficient method to find the likely tumor location [20].

Although asymmetry measures based on histograms make the dissimilarity of healthy parts in image details less pronounced, it can still be more visible than that caused by tumors. The conventional filtering methods may be used for various detections and it is very difficult to achieve a high detection quality due to enormous variations in brain tumor regions, whereas CNNs require enormous computation and data resources. If one wishes to develop a brain tumor detection system of good processing quality at a very low computation cost, the symmetrical nature of brain structure can be explored but the problem of non-tumoral asymmetry should be solved. Also, new methods for feature extraction and classification need to be developed in order not to hit the limits of the existing ones.

## 3. Two-D Gray Level Distribution of 3D data

A tumor can appear in any location in a 3D brain structure and tumor regions can have various gray level distributions. The histograms illustrated in Fig. 2 (a) and (b) are the gray level distributions of two brain images and those of their tumor regions. Such histograms provide us with important statistical characters of a 3D image data, but without locational information about the tumor regions. In this section, we propose a 2D histogram presentation, bringing the locational information to the gray level distributions.

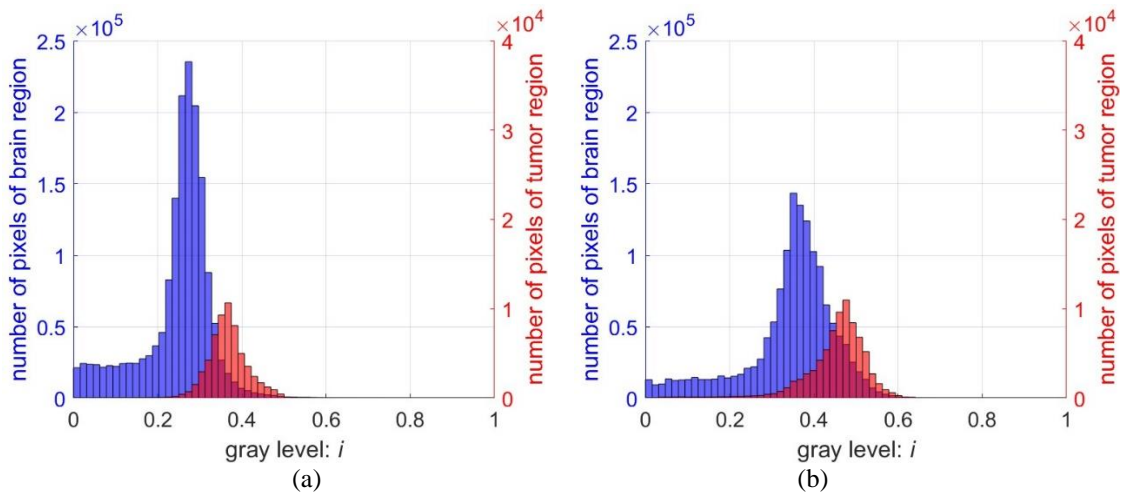

Fig. 2  Gray level distributions of 3D brain images given by the patient cases 01417 and 01438 from BraTS2021 dataset [21]. The pixels outside the brain regions are excluded.

It is known that a 3D brain image can be presented as a series of axial, coronal or sagittal slices, and each slice has a gray level distribution. A 2D histogram presents collectively a series of distributions given by a series of slices, as one example shown in Fig. 3 (a). Let $H(i,j)$ denote such a histogram, the $i$-axis specifies the gray level, normalized to [0, 1], and the $j$-axis is the slice index, i.e., one of the 3 coordinates in the 3D structure.

If $j_0$ is given, $H(i, j_0)$ is the gray level distribution of the pixels in the $j_0^{th}$ slice, whereas if $i = i_0$, $H(i_0,j)$ represents the locational distribution of the pixels at the gray level $i_0$ over the slices in the series. Hence, a 2D histogram $H(i,j)$ encompasses gray level distribution and locational distribution of the pixels.

The 2D histogram illustrated in Fig. 3 (a) is made of the pixels inside the brain region of the 155 slices from a 3D Flair brain image. It demonstrates that a vast majority of the pixels are found (i) in the gray level range (0.2, 0.4) and (ii) in the slices indexed 15 to 142. In other words, the first 14 slices and the last 13 slices (indexed 143 ~ 155) are outside the effective brain region. Hence the coordinates in the y-axis define the location of the brain region in the direction perpendicular to the slices.

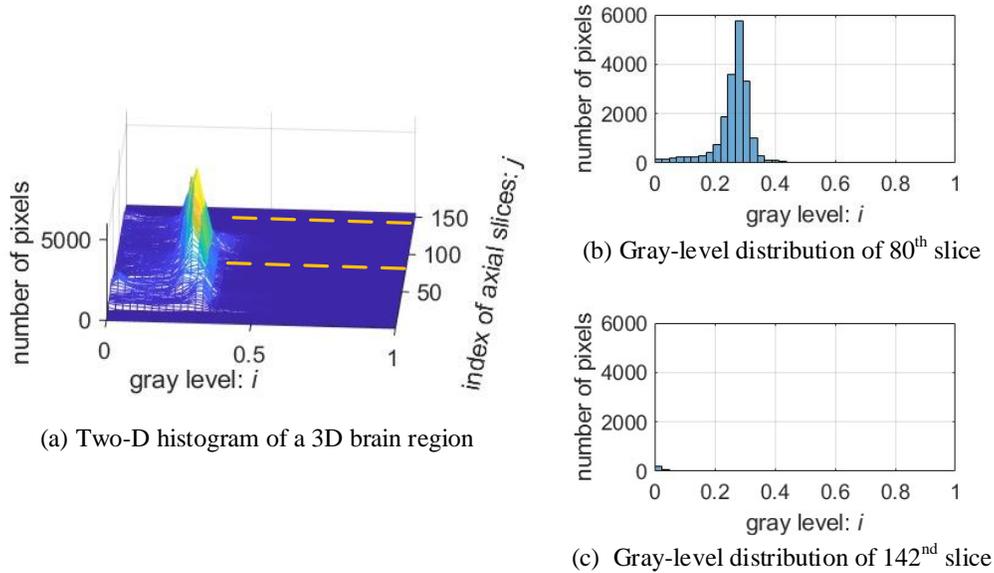

(a) Two-D histogram of a 3D brain region

(b) Gray-level distribution of 80$^{th}$ slice

(c) Gray-level distribution of 142$^{nd}$ slice

Fig. 3  (a) Two-D histogram of the 3D brain region of the patient case 01417 from BraTS2021 dataset. It is composed of a series of gray level distributions of the 155 2D axial slices, excluding the pixels outside the brain region. The x-axis specifies the gray levels, normalized to [0, 1], the y-axis the index of axial slices, the z-axis the number of pixels.
(b) Distribution of the 80$^{th}$ axial slice.
(c) Distribution of the 142$^{nd}$ axial slice.

The 2D histogram shown in Fig. 4 (a) is given by the pixels in a 3D brain region whereas that in Fig. 4 (b) by the tumor pixels, i.e., the pixels in the 3D tumor regions inside the brain. The latter illustrates not only what the gray level distribution of the tumor pixels looks like, but also which slices contain the tumor pixels and which slices are tumor-free. The 2D histogram shown in Fig. 4 (c) and (d) are of another patient case and plotted in the same manner. Comparing the two cases, one can get the following observation.

- The gray level distributions can vary a lot from case to case. The tumor regions in different cases have different brightness.

- The difference in tumor locations in the 2 cases are illustrated in Fig. 4 (b) and (d). Fig. 4 (b) shows the tumor appears in the higher index-numbered section. As the axial slices are index-numbered from the bottom up, the tumor is found in the very top of the brain. The tumor in the 2nd case, shown in Fig. 4 (d), is visibly in the lower section of the brain.

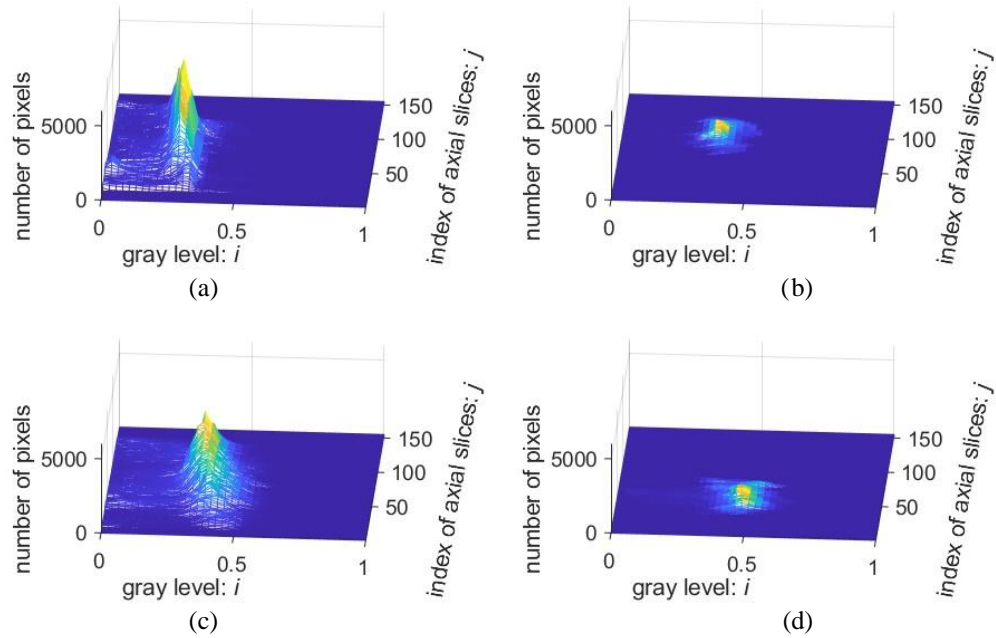

Fig. 4 (a) Two-D histogram of a 3D brain region in a series of axial slices, given by the case 01417 from BraTS2021 dataset.
(b) Two-D histogram of the tumor region (ground truth data).
(c)(d) Two-D histograms given by the case 01438 from BraTS2021 dataset.

If a 3D brain image is sliced 3 times, resulting in axial, coronal, and sagittal slice series, one will have three 2D histograms representing the gray level distributions of the pixels over the 3 series, respectively. Fig. 5 illustrates such a case. The 2D histograms shown in Fig. 5 (b) (d) and (f) represent the gray level distributions of tumor regions in the three series, respectively. They indicate, on one hand, the tumor location in the 3D brain image, and on the other hand, the tumor-free axial, coronal or sagittal slices.

It should, however, be noted that, in a real detection case, 2D histograms of tumor pixels are not available. Nevertheless, they are predictable. We propose a method to use the information from 2D histograms of a brain image to predict the gray level distribution of the tumor pixels inside the image. Based on the results of the prediction, the task of the brain tumor detection can be done easily and effectively to achieve a good processing quality. The procedure of the prediction and the detection is described in the following subsection.

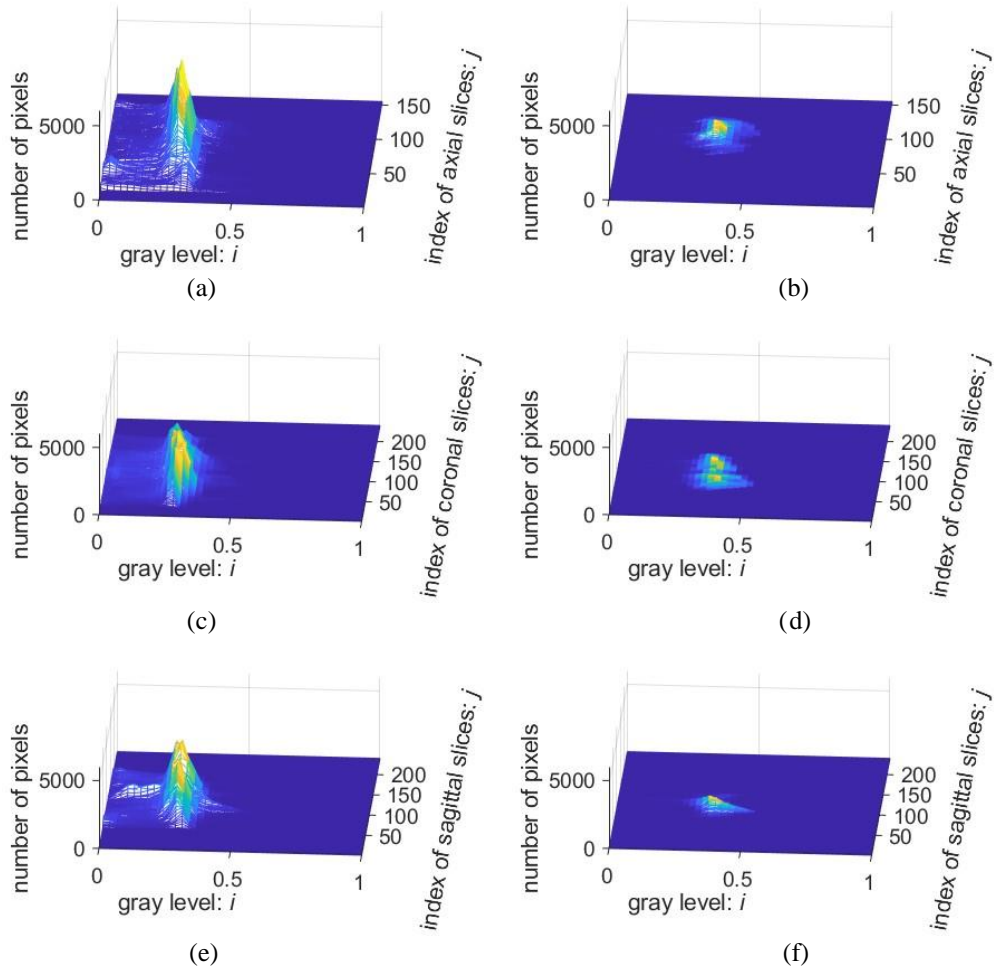

Fig. 5  Two-D histograms of a brain region and that of the tumor region, given by
(a)(b) the axial slices,
(c)(d) the coronal slices, and
(e)(f) the sagittal slices.
The original data sample is from the case 01417 from BraTS2021 dataset.

## 4. Proposed System

The proposed system is designed to predict 2D histograms of the pixels in the tumor region in a 3D brain image of Flair modality, and the prediction results are used to detect brain tumor in pixel-wise precision.

This section is organized as follows. In Subsection 4.1, the overview of the proposed system is presented. In Subsection 4.2, a method to extract tumor information by a particular measure of brain structure asymmetry is described. The data extracted by the asymmetry measure have a significantly higher tumor information density, with respect to that in the input 3D data, but need to be modulated for the prediction. A modulation function is proposed and described in Subsection 4.3. Three-step coarse prediction and the finalization of the tumor pixel distribution are found in Subsections 4.4. Subsection 4.5 is about the brain tumor detection based on the predicted 2D histograms.

## 4.1 System Overview

Of a 3D input image, the object region takes, in general, only a very small percentage of the space and thus the density of the object information is extremely low in the input data. In case of brain tumor detection and there is a thick tumor-free margin in each of the 6 sides of the 3D input. In other words, in each of the 3 series of slices, namely axial, coronal and sagittal series, only a small number of slices contains tumor pixels, and the other slices are tumor-free. However, as a tumor can be found in any place in the 3D brain, it is not easy to localize these slices in the series. Moreover, though we know that the gray levels of the tumor are mainly found in an upper section of the range of the brain region, there is no model relating the gray level distribution of the pixels inside the tumor space to that of the entire brain. Hence, it is very challenging task to predict the 2D histograms with a good precision.

The proposed system is designed to explore 3 commonly known points.

- Though the object location is unknown, some object-free regions can be localized with some certainty. One can identify/remove object-free regions in multiple steps, starting from the most obvious ones, and each step results in a higher density of object information.

- A higher object information density in the input data leads to a better processing quality.

- Since a 3D input image can be sliced three times in the three different directions, i.e., x, y, z axis, resulting in 3 different series of slices, one can design a 3-step process and each step can be performed with a different series of the same 3D data.

The processing scheme in the proposed system is shown in Fig. 6. It has 3 prediction steps interleaved with 3 cropping operations. In each step, the 3D data is sliced in one of the 3 directions, the 2D histogram of the tumor pixels of this series is predicted, and the result is then applied to crop out object-free margins, i.e., tumor-free slices. The cropped 3D data is expected to have a higher object information density, with respect to that in the preceding step, and are then used for the prediction in the following step. In this way, the prediction result can be improved step by step.

The proposed system also involves the generation of asymmetry maps and a modulation function, as shown in Fig. 6. The modulation function is generated from the input data and used to modulate the asymmetry map, or pixel distribution, in each step to produce the prediction result. In the present design, the same modulation function is applied to all the 3 coarse prediction steps, and a modified version to the finalization of the prediction.

The progressive removals of tumor-free regions in the proposed scheme transforms the input 3D image into a 3D minimum bounding box, in which most of the pixels are inside the tumor region. This minimum bounding box is then used to finalize the 3 predicted 2D histograms, as shown in Fig. 6.

The proposed system also includes a simple procedure of brain tumor detection, in which the predicted gray level distribution of the tumor pixels is used to localize them in the minimum bounding box to segment the brain tumor region in a pixelwise precision.

The quality of the final results is related to the data processing quality in each of the prediction steps. In particular, as the operations are performed sequentially, the first prediction and cropping are very critical. The design of the blocks is presented in the following subsections.

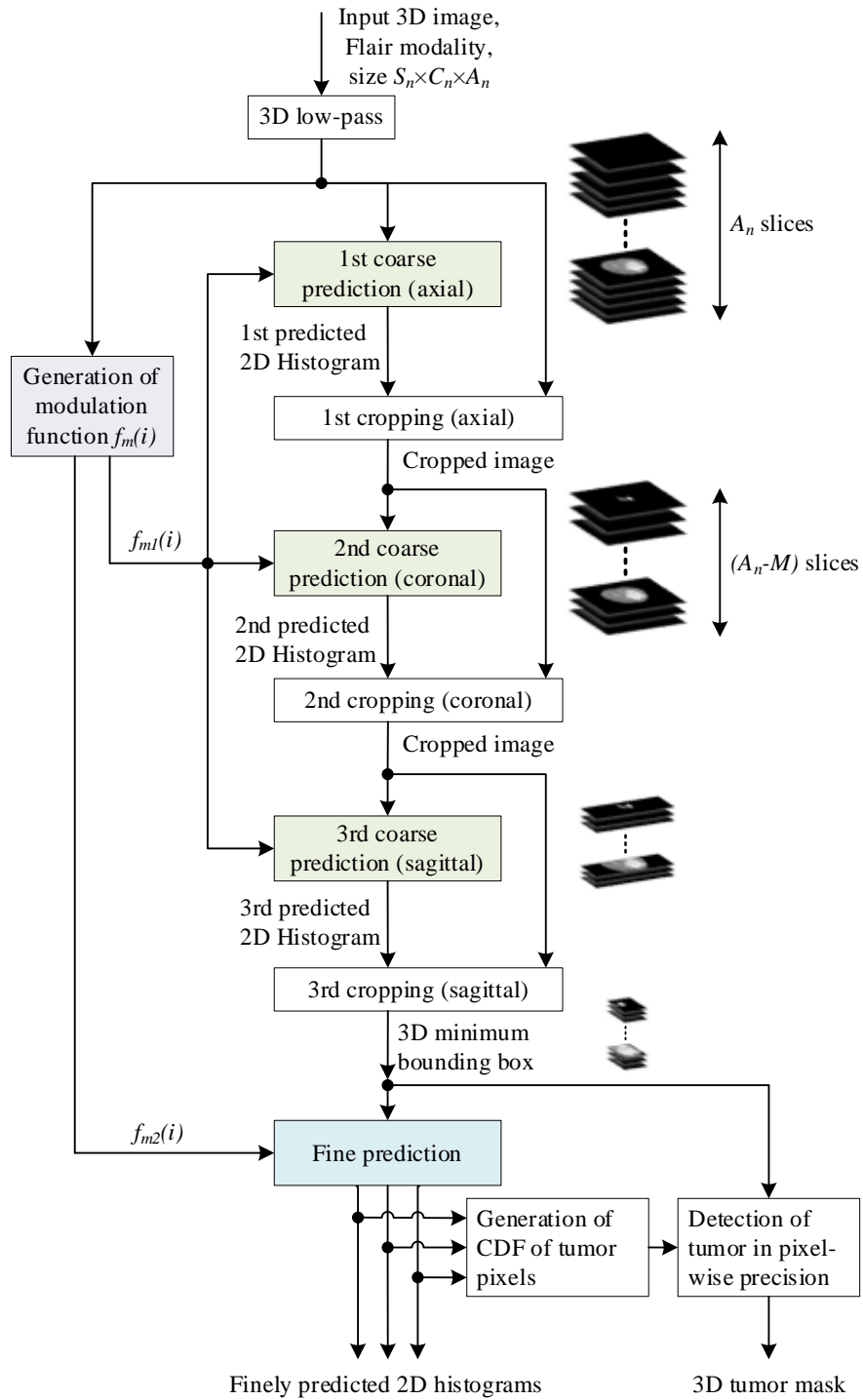

Fig. 6  Block diagram of the proposed system. It receives a 3D brain image input that can be sliced into a series of $A_n$ axial slices, or a series of $C_n$ coronal slices or $S_n$ sagittal slices, and generates the predicted 2D histogram of the brain tumor region for each of the 3 series of slices and a 3D tumor mask.

## 4.2 Brain Image Asymmetry Measure to Extract Tumor Information

In the proposed system, the brain asymmetry is measured for the extraction of the brain tumor information. The results are presented in 2D histograms to indicate how the asymmetry evolves from slice to slice. In this subsection, the details of the measurement are described and the analysis of the data is presented.

A healthy human brain looks left-right symmetrical, though its details are not really left-right mirrored [22]. A tumor-free axial slice shown in Fig. 7 (a) is an example. The presence of a brain tumor causes a more noticeable asymmetry in its structure, as shown in Fig. 7 (b) and (c). Hence, the asymmetry measures in brain images have been used to detect brain tumors [19][20][23]. It should, nevertheless, be noted that, though the tumor-related asymmetry is salient for trained human eyes, it is not prominent in an asymmetry measurement in computer vision. The results of the measurement can be more dominated by the elements representing the natural asymmetry in brain image details than those of the asymmetry caused by tumor, referred to as tumoral asymmetry.

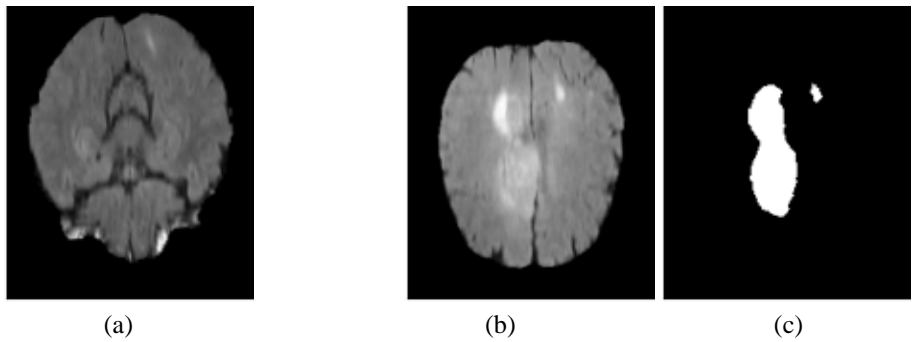

(a) (b) (c)

Fig. 7 (a) Slice of brain image without tumor. The left-right asymmetry in image details is referred to as natural asymmetry.
(b)(c) Slice of brain image with tumor and its binary tumor mask. The asymmetry is more noticeable.

The natural asymmetry in brain images is in image details, reflecting different tissues and fluid, whereas the tumoral asymmetry is more in brain structure. Before all the measures, a 3D low-pass filtering is applied to the input and then each slice is down-sampled to erase some image details so that the elements of natural asymmetry are less dominant in the asymmetry measures.

In the proposed prediction process, the left-right asymmetry of a 3D brain image is measured simply by means of the difference between the 2 histograms given by the left and right halves, respectively, so that the natural asymmetry in fine image patterns is less counted. It should be noted that, in this measure, all the histograms are 2D so that each of them indicates the gray level distributions with the coordinates in one of the 3 dimensions. Let $\Delta H(i,j)$ denote the unsigned histogram difference of the 2 halves, and it is expressed as follows.

$$\Delta H(i,j) = \left| H_{left}(i,j) - H_{right}(i,j) \right| \tag{1}$$

where $H_{left}(i,j)$ is the 2D histogram of the left half and $H_{right}(i,j)$ is that of the right half, $i$ representing the gray level, scaled between 0 and 1, and $j$ the slice index in the axial or coronal series. As the 2D histogram $\Delta H(i,j)$ represents the gray level distribution of the asymmetry elements over the series, it is referred to as asymmetry map.

As the gray level range of a tumor space is in an upper-level section of that of the brain region, the pixels having their gray levels below the mean level of the 3D brain region are not included in *ΔH(i,j)*. In other words, in the asymmetry maps presented in this section, the gray scale is normalized to the range of [0,1] with $i = 0$ corresponding to the mean level of the 3D brain region.

Fig. 8 illustrates an example of $H_{left}(i,j)$, $H_{right}(i,j)$ and $\Delta H(i,j)$ obtained from a 3D Flair image of a typical patient case, in comparison with $H_T(i,j)$, the 2D gray level distribution of the true tumor region, referred to as the ground truth. Comparing $H_{left}(i,j)$ and $H_{right}(i,j)$, one can see the right half has more pixels in the upper gray levels, indicating the presence of a tumor, which is also reflected in $\Delta H(i,j)$. Comparing $\Delta H(i,j)$ and the ground truth $H_T(i,j)$, one can clearly see that the distribution in the upper level range in $\Delta H(i,j)$ is highly correlated to that of $H_T(i,j)$, but that in the lower level range is not.

Evidently, the upper-gray-level section of *ΔH(i,j)* is dominated by the pixels in the tumor region, representing more the tumoral asymmetry. The section of the lower gray levels in *ΔH(i,j)* is, however, more relevant to the natural asymmetry. With a view to obtaining a good prediction of the gray level distribution of the tumor region, the data of *ΔH(i,j)* needs to be modulated so that the elements related to the natural asymmetry will be attenuated. The development of the modulation function is presented in the next subsection.

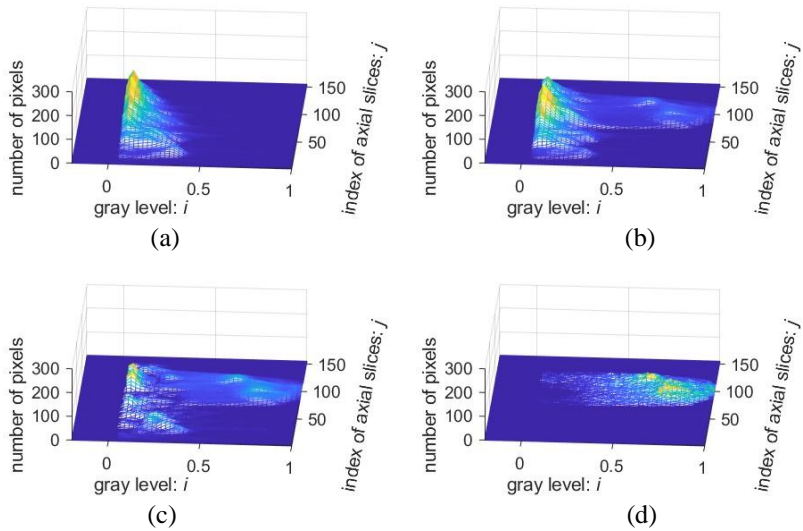

Fig. 8 Four 2-D histograms obtained from the 155 axial slices of a low-passed 3D Flair image sample. The X-axis is the normalized gray scale and the zero point corresponds to the mean value of the 3D brain region, excluding the pixels of gray level values below the mean. The data sample is from the case 01412 of BraTS2021 dataset.
(a) $H_{left}(i,j)$, the 2D histogram of the left half of the 3D image,
(b) $H_{right}(i,j)$, the 2D histogram of the right half of the 3D image,
(c) $\Delta H(i,j) = |H_{left}(i,j) - H_{right}(i,j)|$, and
(d) $H_T(i,j)$, the ground truth of the 2D gray level distribution of the tumor region.

## 4.3 Generation of the Modulation Function

The asymmetry measurement results in a 2D histogram *ΔH(i,j)* representing the natural and tumoral asymmetries in the consecutive slices of a 3D brain image. To generate, from *ΔH(i,j)*, a 2D histogram $H_m(i,j)$ resembling the true gray level distribution of the tumor pixels, one needs to attenuate the elements of natural asymmetry in $\Delta H(i,j)$. As such elements are found in the lower part of the gray level range of $\Delta H(i,j)$, we propose a modulation function $f_m(i)$, of which the characteristic is shown in Fig. 9, and $H_m(i,j)$ will simply be the product of $\Delta H(i,j) \cdot f_m(i)$.

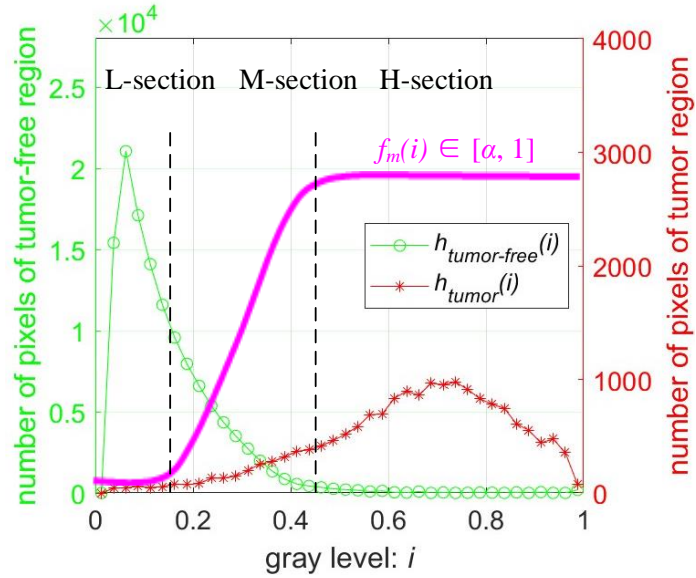

Fig. 9 Graph of the modulation function $f_m(i)$, plotted in magenta, that matches well $h_{tumor-free}(i)$ and $h_{tumor}(i)$, the gray level distribution of the tumor-free region and that of the tumor region, respectively. Ideally, its low (L), mid (M) and high (H) sections should adapt to each individual patient case. The range of $f_m(i)$ is set to be $[\alpha, 1]$ and $\alpha \ll 1$.

Let us divide the gray level range of the modulation function $f_m(i)$ into L-section, M-section and H-section, as shown in Fig. 9. L-section, in which $f_m(i)$ has its lowest value, covers the gray level range where most pixels are in the tumor-free region and H-section covers that of the tumor region, whereas M-section should cover the range where each gray level bin contains pixels of both tumor and tumor-free regions. Ideally, the 3 sections of $f_m(i)$ should match the gray level distribution of the tumor region and that of the tumor-free region in each patient case, but neither of them is available. Hence, the data of the input brain image is the only information source to be used to establish the modulation function $f_m(i)$.

In case of brain tumor detection, the data of the left or right half of a brain is available to generate its 1D and 2D histograms. Since a tumor region usually appears in the left or right half, the half involving the tumor will have its histogram more populated in the upper gray levels than the other half. Let us call the first half *tumor-half* and the other *tumor-free-half*, and $h_{tumor-half}$ and $h_{tumor-free-half}$ denote their 1D histograms, respectively. Fig. 10 (a) and (b) illustrate $h_{tumor-half}$ and $h_{tumor-free-half}$ given by 2 very different patient cases, and each pair is superimposed with the ground truth $h_{tumor-free}$, the normalized gray level distribution of the pixels outside the tumor space in the entire 3D brain region. One can find, in each of these 2 cases, a high degree of similarity between $h_{tumor-free-half}$ and $h_{tumor-free}$. It indicates that, in the half that is less affected by the tumor, the statistical characters of the data are not much different from those of the entire tumor-free regions of the brain. Thus, $h_{tumor-free-half}$ can be used to emulate $h_{tumor-free}$ to determine M-section and H-section of the modulation function $f_m(i)$.

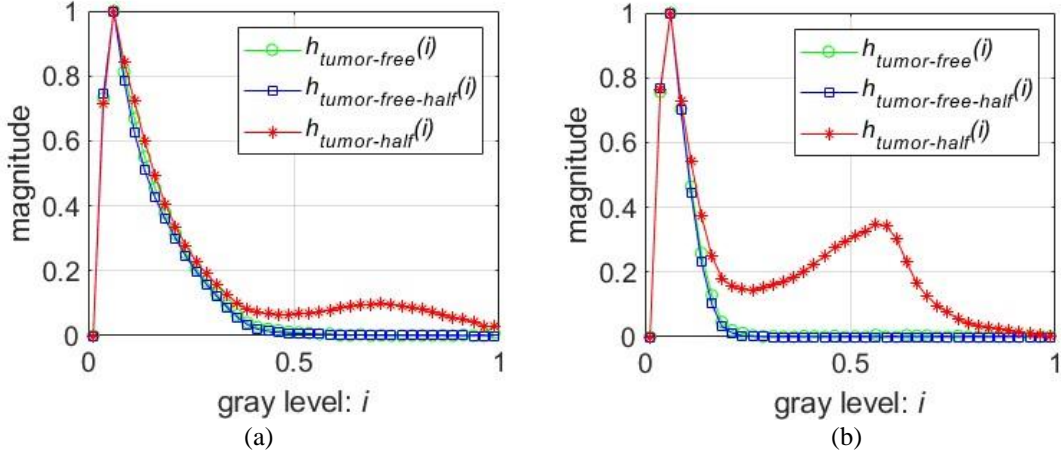

Fig. 10 Gray level distributions given by patient cases 01412 and 01414 from BraTS 2021 dataset. In the 2 graphs, $h_{tumor\text{-}free\text{-}half}$, the distribution of the pixels from the tumor-free half is compared with $h_{tumor\text{-}half}$, that of the tumor half and $h_{tumor\text{-}free}$, that of the true tumor-free region inside the entire brain region.

Of the 1251 patient cases available in BraTS2021 datasets, approximately 94% have tumors developed in either left or right half, and the above observation/analysis is valid for a vast majority of patient cases. Even though a tumor grows in the middle, its region can hardly straddle the left and right halves symmetrically. Hence, the histogram of the half having fewer tumor pixels bears a similitude of $h_{tumor\text{-}free}$ and thus can substitute it to determine $f_m(i)$.

The procedure to generate the modulation function $f_m(i)$ from the distribution of the pixels in the tumor-free half has 2 steps. The first step is to identify which of the 2 halves is more likely to be tumor-free, and the second step is to transform the distribution of the pixels of the identified half into a desired $f_m(i)$.

As the pixels in tumor regions are in the upper-gray-level section, the tumor-free half of the 3D brain image should have a smaller number of high-gray-level pixels with respect to the other half. Hence, the identification is done by simply counting the number of pixels in the upper-gray-level section. In case of samples from BraTS2021, this section is defined as [0.55, 1] in the normalized gray scale, in which the point $i = 0$ corresponds to the mean value of the pixels in the brain region. Let $N_{left}$ denote the number of the pixels in the upper-gray-level section of the left half, and $N_{right}$ that of the right half. Let $h_{tf}(i)$ denote the 1D gray level distribution of the identified half that is presumed tumor-free, and it is calculated as follows.

$$N_{left} = \sum_{i=0.55}^{1}\sum_{j=1}^{N_s} H_{left}(i,j), \ N_{right} = \sum_{i=0.55}^{1}\sum_{j=1}^{N_s} H_{right}(i,j) \quad (2)$$

$$h_{tf}(i) = \begin{cases} \sum_{j=1}^{N_s} H_{left}(i,j), & if \ N_{left} \leq N_{right} \\ \sum_{j=1}^{N_s} H_{right}(i,j), & otherwise \end{cases} \quad (3)$$

where $N_s$ is the number of the slices.

Transforming $h_{tf}(i)$, the distribution of the pixels of the identified half, into a desired $f_m(i)$ is done mainly by truncation and inversion. A block diagram of the transformation presented in Fig. 11 (a), and the curves of the data in this process is visualized in Fig. 11 (b). The 1D histogram $h_{tf}(i)$ of the tumor-free half, plotted in blue, is the input of the process. It is truncated to limit the heights of its bins, resulting in $h_T(i)$ plotted in cyan. The curve of $1/h_T(i)$, plotted in black, can be adjusted to approach the expected $f_m(i)$, plotted in magenta. In this process, low-pass filtering operations are applied before and after the inversion to remove the discontinuity in the curves. The mathematic expressions used in the transformation process are as follows.

$$h_T(i) = \begin{cases} \max_T, & h_{tf}(i) > \max_T \\ h_{tf}(i), & \min_T \leq h_{tf}(i) \leq \max_T \\ \min_T, & h_{tf}(i) < \min_T \end{cases} \quad (4)$$

$$f_m(i) = \left[\frac{1}{h_T(i)}\right]^\gamma + \alpha \quad (5)$$

where $\max_T$ and $\min_T$ are the pre-determined highest and lowest bin-heights, $\gamma$ is a correction factor, and $\alpha$ is a constant.

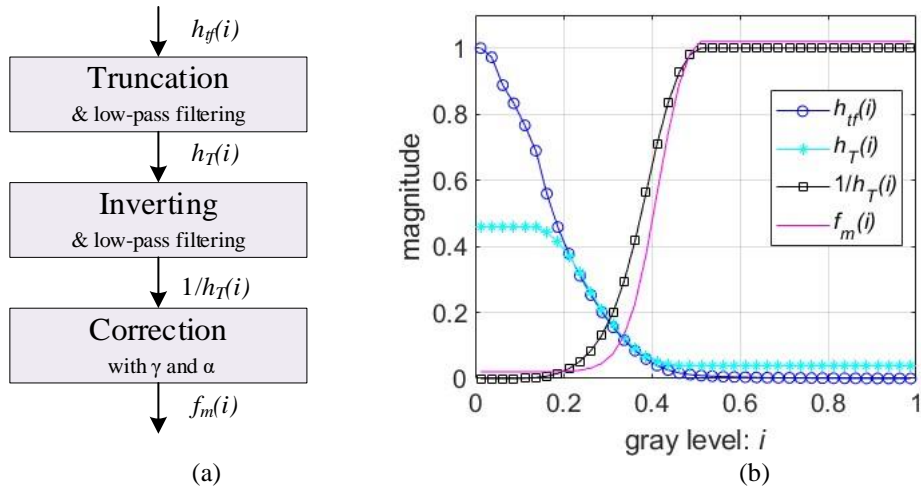

Fig. 11 (a) Block diagram of the procedure to transform $h_{tf}(i)$ to $f_m(i)$.
(b) Graph of $h_{tf}(i)$ of a 3D Flair image, $h_T(i)$, truncated $h_{tf}(i)$ with $\max_T = 0.48$, $\min_T = 0.05$, $1/h_T(i)$ and $f_m(i)$ given by (5) with $\gamma = 1.8$ and $\alpha = 0.02$.

The modulation function $f_m(i)$ can be adjusted by means of the four parameters, $\max_T$ and $\min_T$, $\gamma$, and $\alpha$. One can use $\max_T$ and $\min_T$ to fine-tune, respectively, the 2 particular points where $df_m(i)/di = 1$, and these 2 points define M-section of $f_m(i)$ curve. The parameter $\gamma$ can be used to modify $df_m(i)/di$ in this section, and $\alpha \ll 1$ to maintain a minimum value of $f_m(i)$. For example, increasing the values of $\max_T$ and $\min_T$ shifts the M-section slightly left-wards, making the modulation "milder", i.e., attenuating less the elements in mid gray level range.

Fig. 12 (a) and (b) illustrate the curves of $f_m(i)$ generated from the Flair images of 2 patient cases, respectively. Each of them is superimposed with the ground truth, i.e., the gray level distribution of the tumor-free region and that of the tumor region. One can see that $f_m(i)$ can be made to adapt to the distributions of the tumor and tumor-free pixels in each case, although it is generated independently without them. Its high-value section covers the gray level range of a large majority of tumor pixels, whereas its low-value section covers that of most pixels in the tumor-free region.

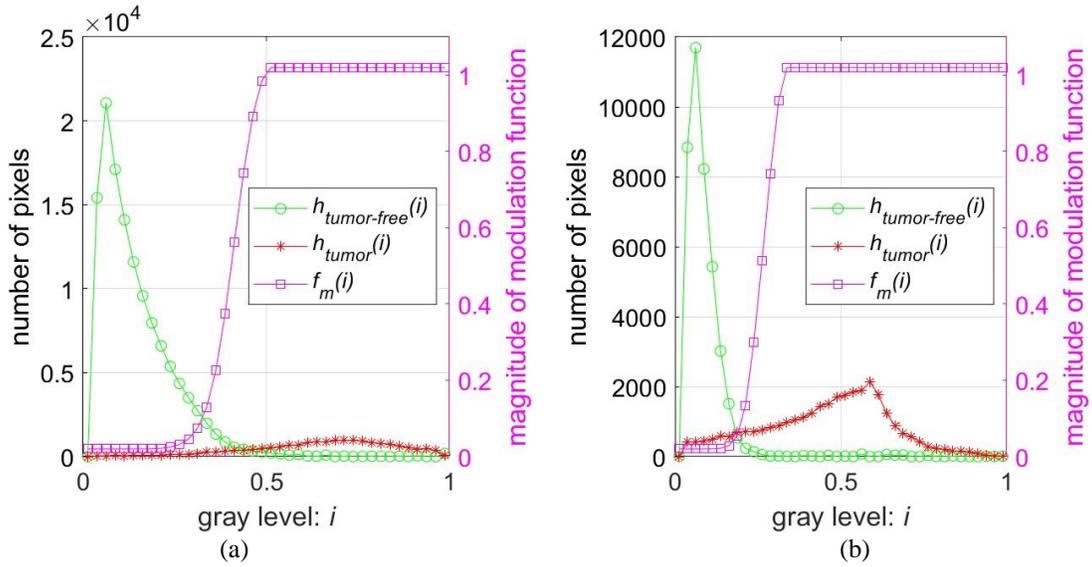

Fig. 12 Two characteristics, plotted in magenta, of the modulation function $f_m(i)$ with $\gamma = 1.8$ and $\alpha = 0.02$, generated with the input data from the Flair modality of the patient cases 01412 and 01414, respectively, of BraTS2021 dataset. In each graph, the curve of $f_m(i)$ is superimposed with $h_{tumor}(i)$, the distribution of the pixels in the tumor region and $h_{tumor-free}(i)$, that of the tumor-free part of brain region. The scale of $f_m(i)$ is [0,1].

## 4.4 Prediction of the Tumor Pixel Distribution

The core of the proposed system is the procedure of 3 coarse predictions of the 2D histograms of the tumor region interleaved with 3 cropping operations, as shown in Fig. 6. The 3 coarse predictions are performed with axial, coronal and sagittal slice series, respectively. The objective of each prediction is to find the concentration of likely tumor pixels in order to identify the likely tumor-free slices that are then removed, i.e., cropped out, from the slice series.

The processing in the first 2 steps are performed with the axial and coronal slices, respectively, as each of them reflects the left-right symmetry of brain structure, allowing to generate an asymmetry map $\Delta H(i,j)$, whereas that in the third step is with a series of cropped sagittal slices. After the 3-step prediction and cropping, the input 3D image is reduced to a minimum bounding box, from which the predicted distribution of the tumor region is refined. The details of the prediction and cropping operations are presented in the following subsubsections.

### 4.4.1 First Two Coarse Predictions and Cropping Operations

In the proposed system, the first coarse prediction is performed on the axial slices. Let $\Delta H_a(i,j)$ denote the asymmetry map generated from the axial slices and $H_{ma}(i,j)$ denote the coarsely predicted distribution of the tumor pixels over the axial slices, we have $H_{ma}(i,j) = \Delta H_a(i,j) \cdot f_{m1}(i)$. The modulation function $f_{m1}(i)$ is defined by (4) and (5), described in Subsection 4.3, generated from the data of the tumor-free half of the original 3D input.

The modulated 2D histogram $H_{ma}(i,j)$ represents the asymmetry in the upper gray levels, where most of the pixels in the tumor region are found. Thus, it is highly correlated with $H_{Ta}(i,j)$, the 2D histogram of the pixels in the tumor region given by the ground truth.

Fig. 13 (a) and (b) illustrates an example of the first prediction results, in which $\Delta H_a(i,j)$ and $H_{ma}(i,j)$ are obtained from the same patient case shown in Fig. 8. Comparing $H_{ma}(i,j)$ with $H_{Ta}(i,j)$ shown in Fig. 13 (c), one can observe that i) $H_{ma}(i,j)$ emulates well the distribution of most pixels in the tumor region and ii) it indicates a slice-index range, very similar to that in $H_{Ta}(i,j)$, where the tumor pixels are located. The same degree of similarity is also observed in the prediction results of a vast majority of the 1251 patient cases in BraTS2021 dataset. Thus, in the absence of $H_{Ta}(i,j)$, $H_{ma}(i,j)$ can be considered as a coarsely predicted gray level distribution of the tumor pixels in the consecutive axial slices.

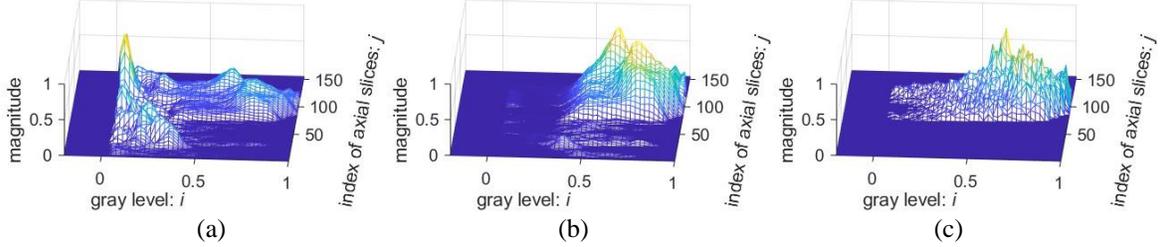

Fig. 13 (a) Asymmetry map $\Delta H_a(i,j)$, generated with the axial slice series of the patient case 01412 from BraTS2021 dataset.
(b) Modulated 2D histogram of the axial slice series, $H_{ma}(i,j)$, to be used as a coarsely predicted 2D histogram of the tumor region.
(c) Two-D histogram of the pixels in the tumor region given by the ground truth, $H_{Ta}(i,j)$, over the axial slices.

To identify the tumor-free axial slices, the $H_{ma}(i,j)$, obtained from the axial series, is transformed into 1D histogram $h_{la}(j)$ to represent the locational distribution of the tumor pixels in the axial slice, which is done as follows.

$$h_{la}(j) = \sum_i H_{ma}(i,j) \qquad (6)$$

where $j = 1 \sim N_s$ and $N_s$ is the number of slices in the series. An example of $h_{la}(j)$ shown in Fig. 14 (b) is obtained from $H_{ma}(i,j)$ shown in Fig. 14 (a). High magnitudes in $h_{la}(j)$ indicate the concentration of pixels of interest, i.e., tumor pixels, in the corresponding slices. The index range of these slices is determined by the two local minima in $h_{la}(j)$ curve. The slices indexed between them are considered slices with tumor, and the others tumor-free. One can see in Fig. 14 (b) that the set of the axial slices identified as tumor slices are almost identical to that in the ground truth. The tumor-free slices, found on the top and bottom of the input 3D image, constitute two tumor-free margins and are then effectively cropped out.

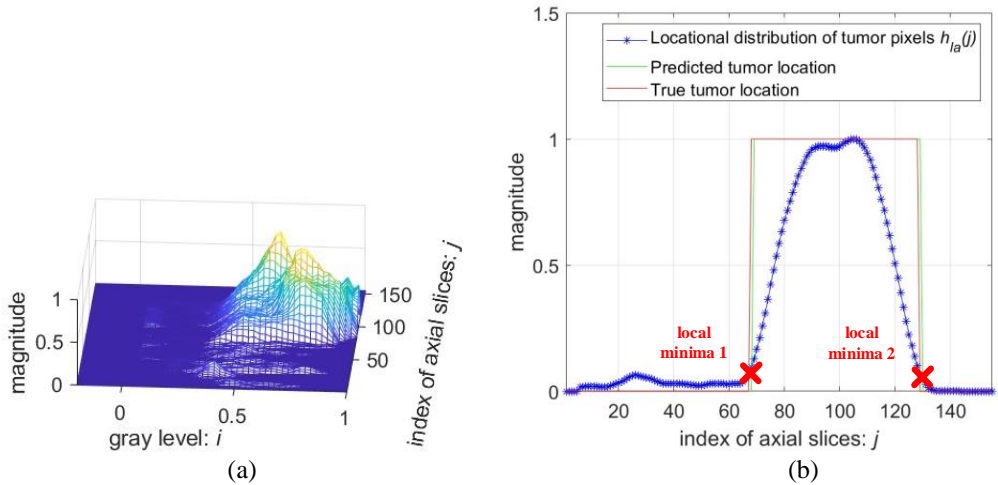

Fig. 14 (a) Coarsely predicted 2D histogram of the tumor region, $H_{ma}(i,j)$, over the axial slices.
(b) Predicted locational distribution of the tumor pixels over the series of axial slices, $h_{la}(j)$, plotted in blue. The 2 local minima define the 2 boundaries of the predicted set of consecutive tumor slices, specified by the green frame, in comparison with the ground truth framed in red.
The data sample is from the patient case 01412 of BraTS2021.

By the cropping operation, the size of the 3D image is reduced significantly, whereas the loss of the tumor pixels is insignificant. Of the 1251 patient cases in BraTS2021, it results in a removal of more than 60% slices from the 3D brain region, while losing less than 4% of the tumor pixels.

The second coarse prediction is then applied to the cropped 3D image presented as a series of coronal slices. The number of slices in this series is the same as that of the original Flair image, but the number of non-zero pixels per slice is much smaller because the predicted tumor-free top and bottom margins of the input 3D image have been cropped out in the first step of prediction/cropping, as examples shown in Fig. 15. Nevertheless, the overall left-right symmetry is preserved in the cropped coronal slices that are not affected by tumor, and a coronal asymmetry map $\Delta H_c(i,j)$ can be calculated to represent the distribution of asymmetrical elements from the cropped coronal slices.

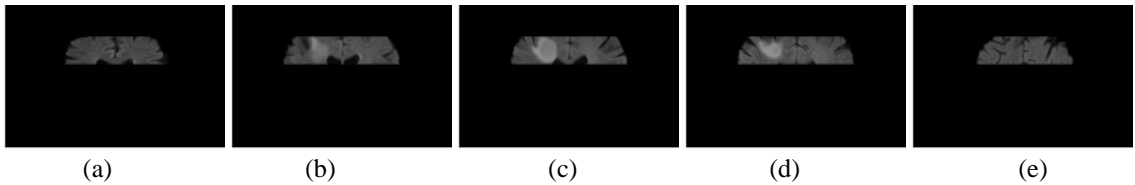

(a) (b) (c) (d) (e)

Fig. 15 Coronal slices sampled from a 3D image, after the cropping operation to remove the tumor-free axial slices.
(a)(e) Cropped coronal slice without tumor. The left half is somehow symmetrical to the right half.
(b)(c)(d) Those with tumor. The left-right symmetry is much less than that in (a) or (e).

The procedure of the second prediction is the same as the first one. The same modulation function $f_{m1}(i)$ is applied to the coronal asymmetry map $\Delta H_c(i,j)$ obtained from the coronal slices to generate the second predicted tumor pixel distribution $H_{mc}(i,j)$. Fig. 16 illustrates the predicted distribution over the coronal slices, in comparison with the ground truth.

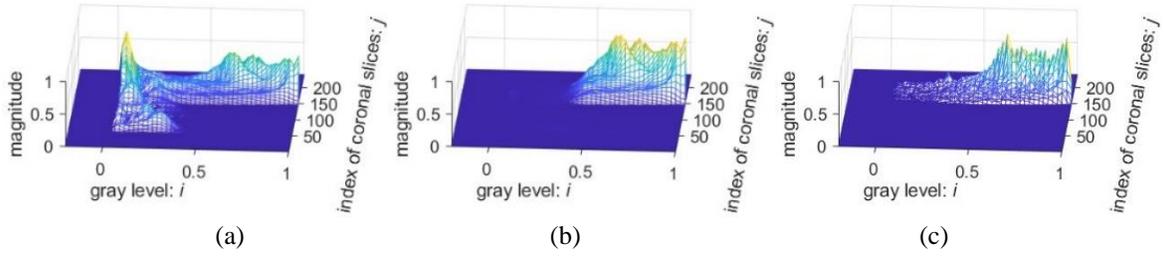

(a)　　　　　　　　　　　　(b)　　　　　　　　　　　　(c)

Fig. 16 (a) Asymmetry map $\Delta Hc(i,j)$, generated with the cropped coronal slice series of the patient case 01412 in BraTS2021.
(b) Modulated 2D histogram of the coronal slice series, $H_{mc}(i,j)$, to be used as a coarsely predicted 2D histogram of the tumor region.
(c) Two-D histogram of the pixels in the tumor region given by the ground truth, $H_{Tc}(i,j)$, over the coronal slices.

The cropping operation following the second coarse prediction is identical to that in the first step. It results in the removal of 2 sets of coronal slices that are considered tumor-free.

After the 2 cropping operations in the first 2 steps, the predicted tumor-free margins in the top, bottom, back and front sides of the original 3D input have been removed. The result of these removals can be seen in sagittal slices. Fig. 17 illustrates a few examples of sagittal slices cropped twice, in comparison with the original ones. The series of cropped sagittal slices is then ready for the next step of prediction and cropping.

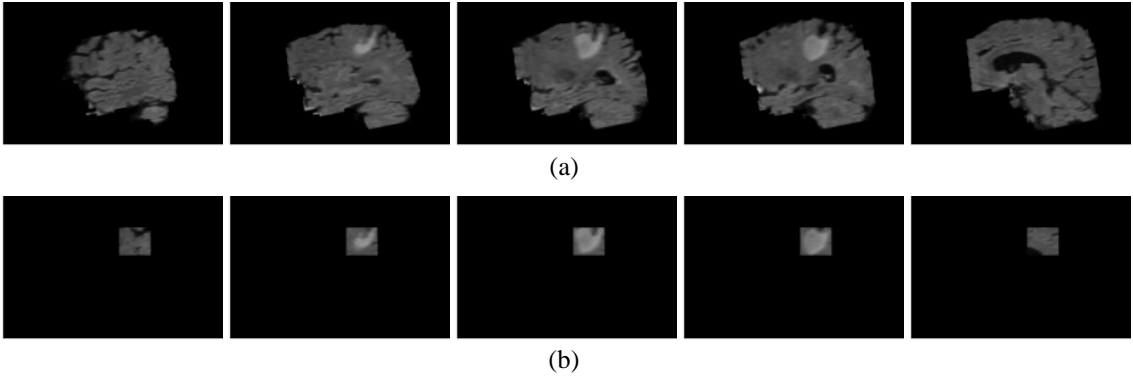

(a)

(b)

Fig. 17 (a) Sagittal slices sampled from an original series, before the 2 cropping operations.
(b) Sagittal slices after the cropping operations applied to the axial and coronal slices. The first and the last sagittal slices are tumor-free, whereas the other 3 involve tumor regions.

### 4.4.2 Third Coarse Prediction & Cropping and the Final Prediction

The objective of the 3rd coarse prediction and cropping is to identify tumor-free sagittal slices and to remove them. As a sagittal slice does not feature left-right symmetry, no asymmetry map can be generated in this step. It should, however, be noted that the percentage of tumor pixels in this sagittal slice series is evidently much higher than that of the original 3D input. In particular, most pixels in the upper gray levels are found in the tumor region. The prediction in this 3rd step is done by modulating $H_s(i,j)$, the 2D histogram of the cropped sagittal slices. The coarsely predicted 2D histogram of the tumor pixels over the sagittal slices is denoted as $H_{ms}(i,j) = H_s(i,j) \cdot f_{m1}(i)$, with the same $f_{m1}(i)$ used in the 2 previous coarse prediction steps. This modulation attenuates the elements in the lower-gray-level section, resulting in a coarsely predicted 2D histogram of the tumor pixels in the sagittal slices, as an example shown in Fig. 18.

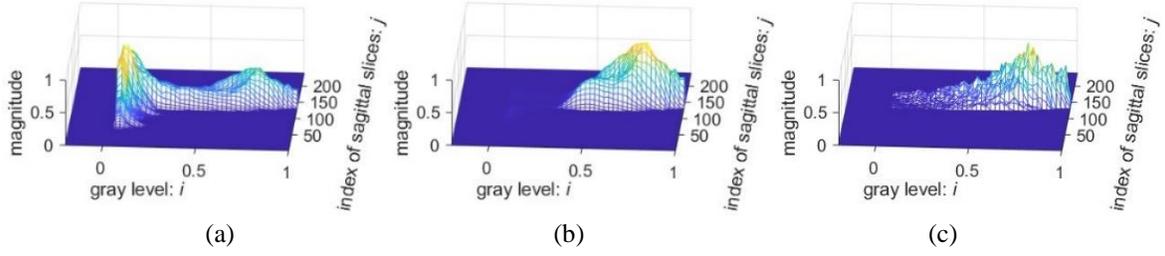

(a)           (b)           (c)

Fig. 18 (a) Two-D histogram $H_s(i,j)$, given by the pixels in the cropped sagittal slice series of the patient case 01412 in BraTS2021.
(b) Modulated 2D histogram of the sagittal slice series, $H_{ms}(i,j)$, to be used as a coarsely predicted 2D histogram of the tumor region.
(c) Two-D histogram of the pixels in the tumor region given by the ground truth, $H_{Ts}(i,j)$, over the sagittal slices.

The cropping operation in this step is identical to that in the other steps. It results in removal of the 2 sets of tumor-free sagittal slices, i.e., the tumor-free margins in the left and right sides of the 3D input. By the 3-step cropping, the original 3D input is reduced to a 3D minimum bounding box, in which most of the pixels are found in the tumor region.

The operation of the fine prediction is applied to the data of the 3D minimum bounding box. Its axial, coronal and sagittal slice series give three 2D histograms, denoted by $H_{ba}(i,j)$, $H_{bc}(i,j)$ and $H_{bs}(i,j)$, respectively. An example is illustrated in Fig. 19 (a) (d) and (g). Comparing the 3 histograms with the ground truth of the tumor pixel distributions illustrated in Fig. 19 (c) (f) and (i), one can notice that i) the 2 sets are very similar in the upper-gray-level section and ii) their differences are found in the lower and mid gray level sections, as the minimum bounding box involves tumor-free regions. Hence, like the coarse predictions, the operation for the fine prediction is to attenuate the elements of $H_{ba}(i,j)$, $H_{bc}(i,j)$ and $H_{bs}(i,j)$ in the mid and lower-gray-level sections by a simple modulation.

It should, however, be noticed that the gray levels of the pixels in tumor regions can cover a wide range. Though a majority of the tumor pixels is found in the upper-gray-level section, a non-negligible minority is found in the middle and lower sections, which should be taken into consideration in the fine prediction. Hence, the modulation function in this stage is adjusted to attenuate less elements in the mid and lower-gray-level ranges, with respect to that in the coarse prediction steps.

The final prediction results in the three 2D histograms, denoted as $H_{pa}(i,j)$, $H_{pc}(i,j)$ and $H_{ps}(i,j)$, indicating the gray level distribution of the tumor pixels over the axial, coronal, and sagittal slice series, respectively. They are expressed as follows.

$$\begin{cases} H_{pa}(i,j) = H_{ba}(i,j) \cdot f_{m2}(i) \\ H_{pc}(i,j) = H_{bc}(i,j) \cdot f_{m2}(i) \\ H_{ps}(i,j) = H_{bs}(i,j) \cdot f_{m2}(i) \end{cases} \quad (7)$$

where $f_{m2}(i)$ is defined by (4) and (5). Compared to $f_{m1}(i)$, the characteristic of $f_{m2}$ is slightly left-wards shifted, the slope, $df_{m2}(i)/di$, in M-section is gentler, and the minimum value of $f_{m2}(i)$ is increased to preserve more pixels in the low-and-mid gray level range. It is done by i) slightly increasing $max_T$ and $min_T$, ii) reducing $\gamma$ and iii) increasing $\alpha$.

An example of the 3 predicted 2D histograms are presented in Fig. 19 (b)(e)(h). The ground truth data of the tumor distributions are presented in Fig. 19 (c)(f)(i). One can find that, the predicted tumor distributions are very similar to the ground truth. They can be used, for example, to detect the whole tumor regions in pixel-wise precision in the brain image.

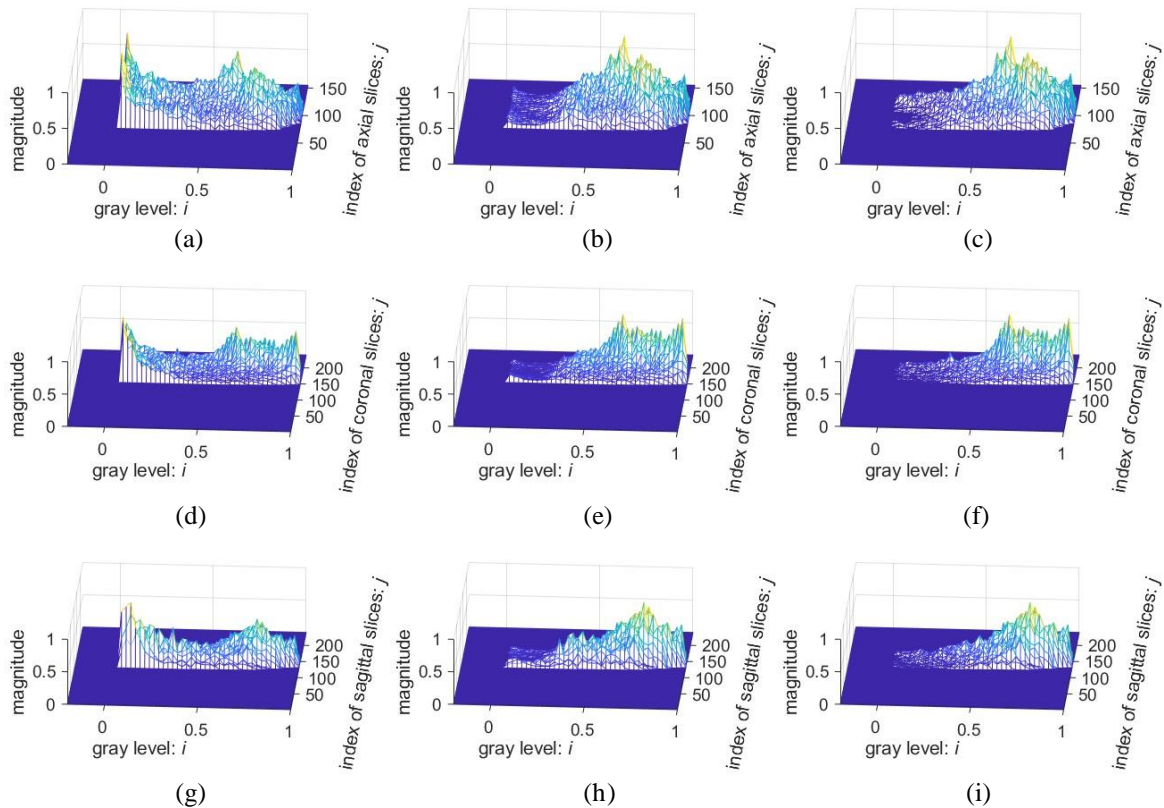

Fig. 19 (a)(d)(g) Two-D histograms of minimum bounding box of axial, coronal and sagittal slice series.
(b)(e)(h) Predicted 2D histograms of tumor pixels in axial, coronal and sagittal slice series.
(c)(f)(i) Ground truth of the 2D histograms in (b)(e)(h).
The data sample is from the patient case 01412 in BraTS2021.

## 4.5 Brain Tumor Detection in Pixel-Wise Precision

In the proposed system, the input data of the brain tumor detection block is a 3D bounding box after the 6 tumor-free margins are cropped out from the original 3D brain image of Flair modality. The gray level distribution of the tumor pixels has been predicted, but the locations of the pixels in this bounding box are not specified. The process in this detection block is to transform the bounding box into a 3D binary tumor mask with pixel-wise precision. The transformation is done by 2 very simple operations, i.e., pixel binarization by gray level thresholding and morphological processing by low-pass filtering.

The binarization is to divide, coarsely by a simple gray level threshold, the pixels in the bounding box into 2 groups, those inside the tumor region and those outside. The threshold should be determined with 2 issues taken into consideration.

- It should be variable to adapt to the gray level distribution of the pixels in individual cases.
- The gray level range of the pixels in the tumor region can extend to a very low point, as the example shown in Fig. 20 (a). If the threshold is determined in such a way that most of the tumor-free pixels are put in one side and most of tumor pixels on the other side, as the graphs in Fig. 20 (a) shows, a small portion of the true tumor pixels will unavoidably be misplaced.

In practice, in order to separate a vast majority of tumor-free pixels from the tumor pixels, it is reasonable to allows, e.g., 20% of tumor pixels to be misplaced if their misplacement is insignificant enough to be corrected in the following processing. In this case, the threshold is found at the gray level point corresponding to 20% in the cumulative distribution function (CDF) of the tumor pixel population, as shown in Fig. 20 (c). The thresholds defined in this manner can be adaptive to the various distributions of individual cases. Since the true distribution of the tumor area is not available, the predicted one is used to determine the threshold, as Fig. 20 (c) shows.

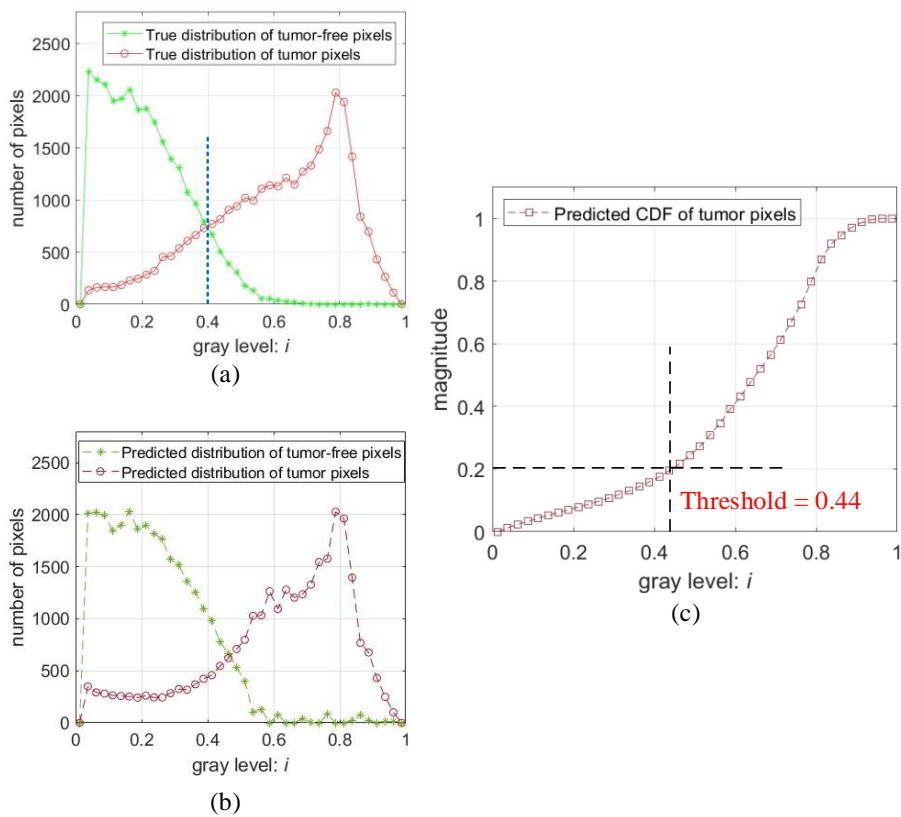

Fig. 20 (a) True gray level distributions of the pixels inside the tumor region, plotted in red, and that of outside the tumor region, plotted in green, of a 3D minimum bounding box. The blue dashed line indicates an assumed gray level threshold. The data sample is from the case 01452 of BraTS2021 dataset.
(b) Predicted gray level distributions, obtained from the original 3D Flair image of the same case.
(c) CDF derived from the predicted distribution of the tumor pixels shown in (b). The gray level threshold is 0.44, corresponding to the level of CDF = 20%.

The binarization by means of a simple thresholding results in a coarse binary tumor mask with a minority of tumor pixels misplaced in the tumor-free group and vice versa. A slice of such a 3D mask is illustrated in Fig. 21 (b). To correct the misplacement, a morphological operation is applied. In this design, it is done by (i) a convolution with a simple 3D averaging kernel of 5×5×5 pixels and (ii) assigning the logic-1 value to all the pixels having their gray levels grater than a pre-determined floor and logic-0 to the others. Fig. 21 (c) illustrates the slice generated by such a morphological operation, in comparison with the ground truth illustrated in Fig. 21 (d).

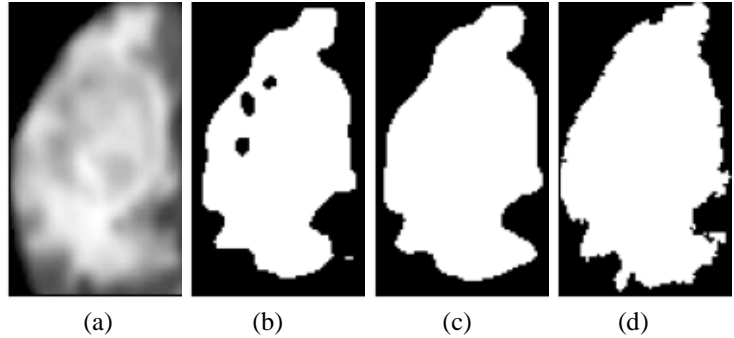

(a)      (b)      (c)      (d)

Fig. 21 (a) Slice from a 3D minimum bounding box generated from the data of the patient case 01418 of BraTS2021 dataset.
    (b) Slice of the coarse mask after the binarization.
    (c) Slice of the final binary mask after the morphological operation.
    (d) Slice of the true tumor mask.

This brain tumor detection is done by the 2 very simple operations, as it is performed on the data of the predicted 3D minimum bounding boxes of tumors and the predicted tumor pixel distribution. Hence it is an application of the prediction results. The quality of the detection depends very much on the quality of the prediction. Various quality measures have been conducted to evaluate the performance of the proposed system. The results are presented in the next section.

## 5. Performance Evaluation

The performance of the proposed system has been evaluated with the available patient cases in BraTS datasets. The quality of the prediction of the tumor pixel distribution in a 3D brain image is assessed. So is the quality of the brain tumor detection. The information about the datasets is found in Subsection 5.1. The prediction and detection results are visualized in Subsection 5.2. The quantitative performance measurements and ablation study are found in Subsection 5.3, and the performance comparison is presented in Subsection 5.4.

### 5.1 Dataset

The processing quality of the proposed system has been measured with the data of BraTS2021 (Brain Tumor Segmentation 2021) [21]. There are 1251 patient cases of MRI scanning and each is accompanied by a ground-truth tumor mask approved by medical specialists. As the system does not need training, the data of all the 1251 patient cases have been used to measure the processing quality of both the prediction of the tumor pixel distributions and the brain tumor detection.

In BraTS2021 dataset there are additional 219 patient cases, referred to as the validation samples, of which the ground truth data is not accessible for public. They have also been used to evaluate the tumor detection quality of the proposed system, by means of the online platform Synapse [24] where the assessment is a standard process with data from the Cancer Imaging Archive [25][26][27][28].

In order to compare the performance of the proposed system, in terms of brain tumor detection, with those published in recent years, earlier versions of BraTS datasets, namely BraTS2018 [29], BraTS2019 [30] and BraTS2020 [31] have also been used for the evaluation. In case of testing on the validation samples of these datasets, the tumor detection results have been evaluated by the online platform, Center for Biomedical Image Computing and Analytics Image Processing Portal (CBICA IPP) [32].

## 5.2 Observation of Prediction and Detection Results

The first step of the performance assessment of the proposed system is to observe (i) the predicted histograms and (ii) the detected 3D tumor masks, in comparison with the ground truth of the datasets. The histograms include three 2D gray level distributions of tumor pixels in the axial, coronal and sagittal series, respectively.

The prediction results of 2 patient cases are illustrated in Figs. 22 and 23. The distributions of tumor pixels in these cases are very different, as shown in Figs. 22 (d)(e)(f) and 23 (d)(e)(f), which is very common in practice. One can see that the predicted histograms are highly similar to the true ones, demonstrating that the proposed system is able to predict the distributions of different cases. The predicted 1D histograms are derived from the 2D ones and are also highly similar to the true ones, as shown in Figs. 22 (g)(h) and 23 (g)(h).

In Fig. 24, four examples of tumor detection results generated by the proposed system are illustrated. The gray level ranges and variations inside and outside the tumor regions are very different in these cases. As described previously, the detection process is a very simple thresholding operation followed by a low-pass-based morphological operation. Finding an appropriate threshold for each individual case is the key to achieve a good detection. A good prediction of the tumor pixel distribution results in a suitable threshold. The 4 detected tumor masks presented in the central column are very similar to the true ones shown the right column, demonstrating that the proposed method is effective to detect varieties of tumors. Though the detected masks look like a bit over-smoothed, compared to the true ones, they can be used to localize the brain tumors with a good accuracy.

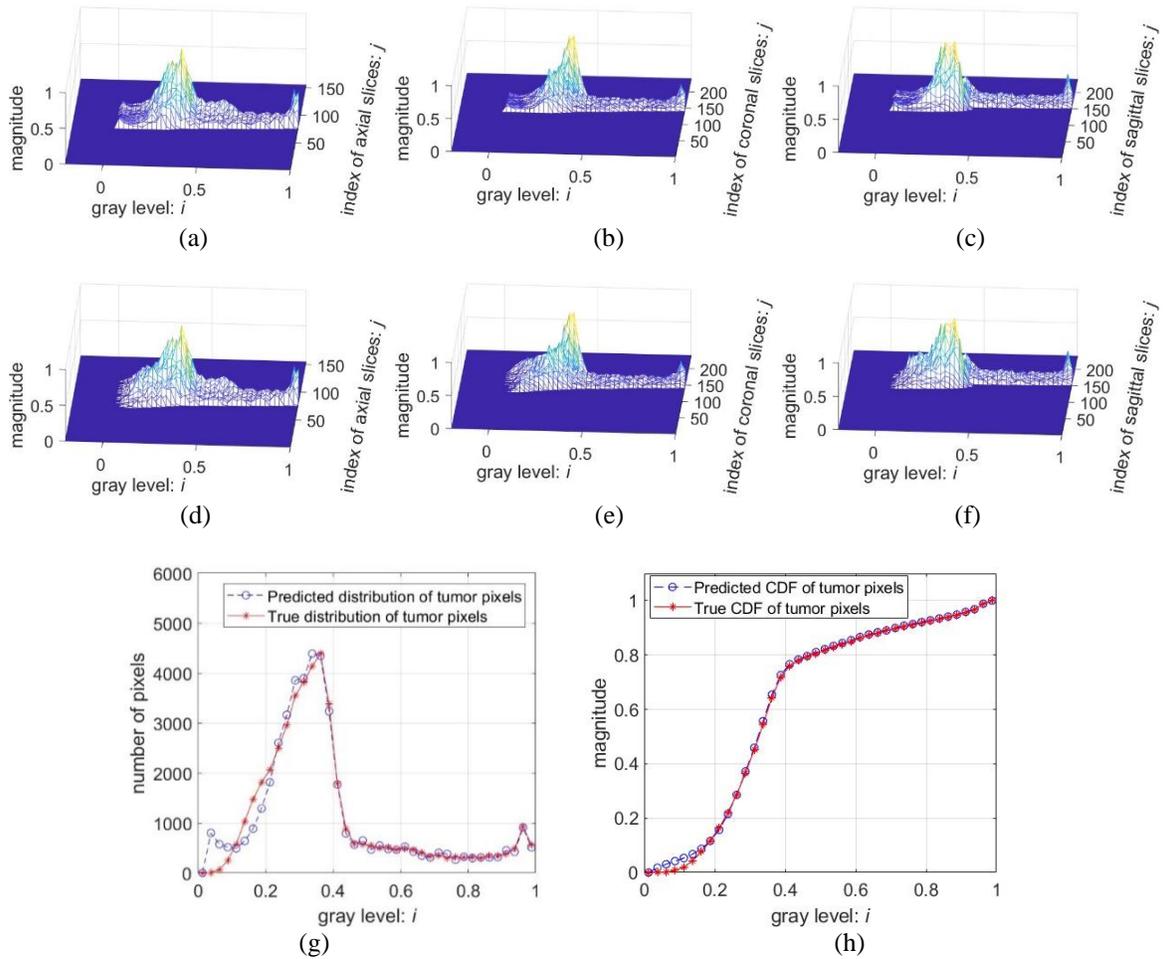

Fig. 22 (a)(b)(c) Predicted 2D gray level distributions of tumor pixels of axial, coronal and sagittal slice series.
   (d)(e)(f) True 2D gray level distributions of the case.
   (g) Predicted and true 1D gray level distribution of tumor pixels.
   (h) Predicted and true CDF of (g).
   The data sample is from patient case 01268 of BraTS2021 dataset.

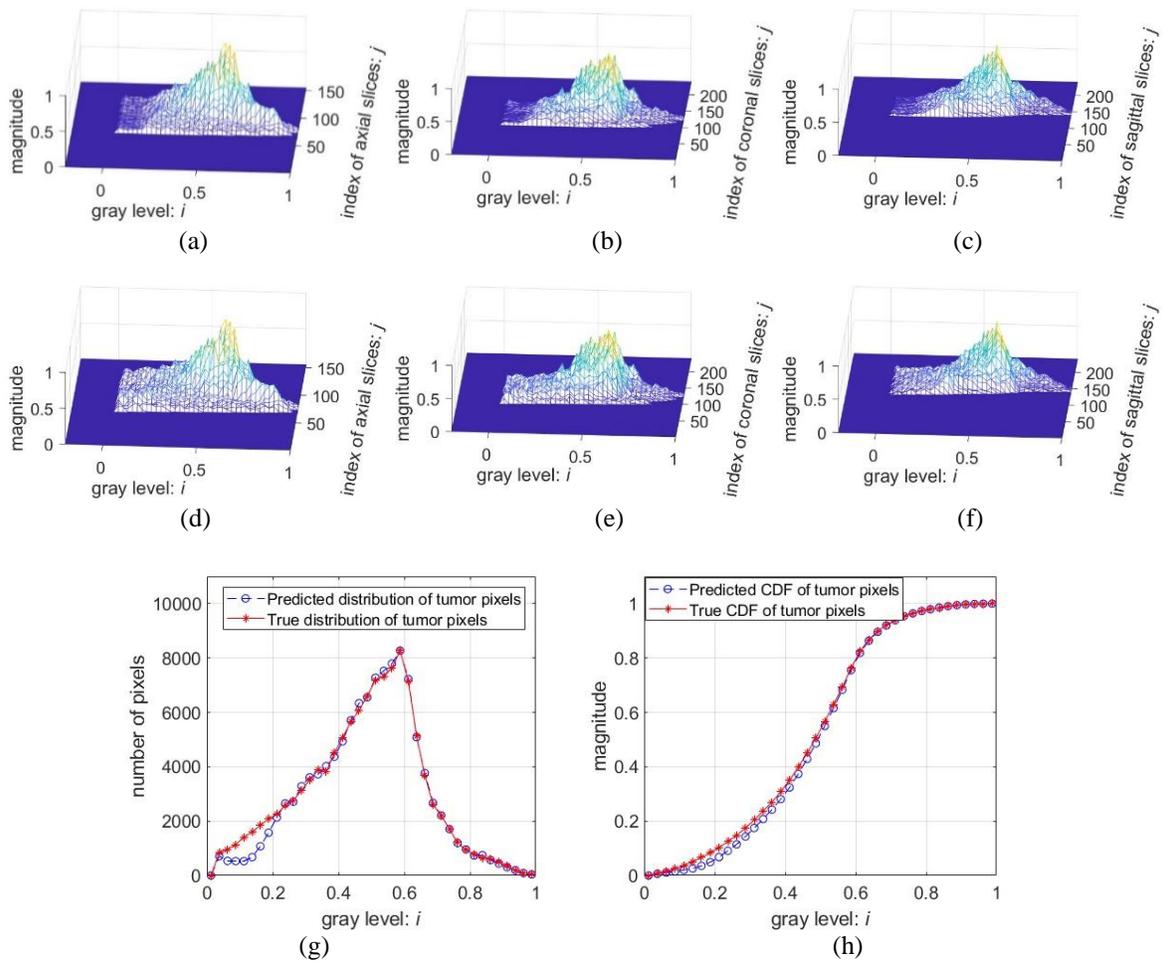

Fig. 23 (a)(b)(c) Predicted 2D gray level distributions of tumor pixels of axial, coronal and sagittal slice series.
(d)(e)(f) True 2D gray level distributions of the case.
(g) Predicted and true 1D gray level distribution of tumor pixels.
(h) Predicted and true CDF of (g).
The data sample is from patient case 01414 of BraTS2021 dataset.

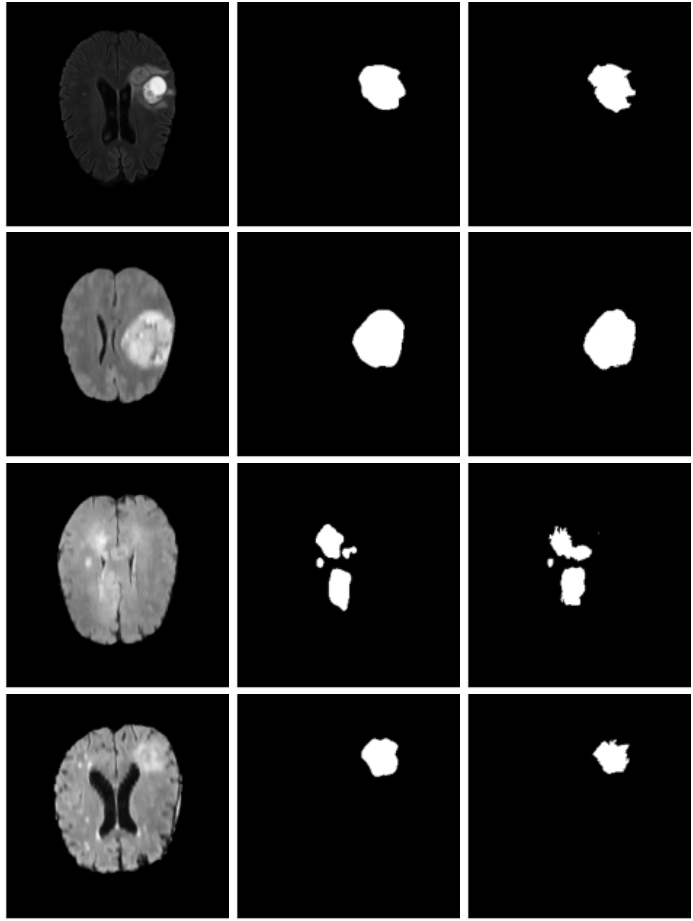

Fig. 24 Four examples of brain tumor detection by the proposed system. The 4 original MRI slices of Flair modality are found in the left column, the 4 detected binary tumor masks in the middle column and the true ones in the right column. The original data are from patient cases 01268, 01414, 01417, and 01420 of BraTS2021 dataset.

Besides the observation, quantitative measurements have been performed on a very large number of samples to assess more objectively the performance of the proposed system. The details are presented in the following subsection.

## 5.3 Quantitative Performance Measurements

The quantitative measurements have been done mainly with the data of the 1251 patient cases of MRI scanning, including the ground truth data, in BraTS2021 dataset. In this subsection, the description of the performance metrics is found in Subsubsection 5.3.1, an ablation study and its result in Subsubsection 5.3.2, and the results of the comprehensive tests of the prediction and detection in Subsubsection 5.3.3 and Subsubsection 5.3.4, respectively.

### 5.3.1 Performance Metrics

As the proposed system is designed to predict the gray level distributions of tumor pixels and to detect tumors, the performance metrics involve 2 kinds of measures, for the prediction and the detection, respectively.

The prediction quality can be measured by the degree of similarity between the predicted and true histograms. Correlation coefficient (CC), mean squared error (MSE) and structural similarity index measure (SSIM) [33] are commonly used for similarity measurement. SSIM is defined as follows.

$$\text{SSIM}(x, y) = [l(x, y)]^\alpha \cdot [c(x, y)]^\beta \cdot [s(x, y)]^\gamma \tag{8}$$

where

$$l(x, y) = \frac{2\mu_x\mu_y + C_1}{\mu_x^2 + \mu_y^2 + C_1}, \quad c(x, y) = \frac{2\sigma_x\sigma_y + C_2}{\sigma_x^2 + \sigma_y^2 + C_2}, \quad s(x, y) = \frac{\sigma_{xy} + C_3}{\sigma_x\sigma_y + C_3}$$

$(x, y)$ are two sets of data, $(\mu_x, \mu_y)$ denote the mean values of $x$ and $y$, $(\sigma_x, \sigma_y)$ are the standard deviations, $\sigma_{xy}$ is the correlation coefficient, $(\alpha, \beta, \gamma)$ are set to be $(1, 1, 1)$, and $(C_1, C_2, C_3)$ are small non-zero constants to stabilize the division with weak denominator.

The detection quality can be measured in Dice coefficient indicating how much a predicted object mask and the true object mask are overlapped. It is defined as

$$Dice(X, Y) = \frac{2TP}{(TP + FN) + (TP + FP)} \tag{9}$$

where $TP$ (true positive) is the overlapped part of the predicted and true object masks, $FN$ (false negative) is the part of the true object mask that is not covered by the predicted mask. The entire true object mask is represented by $(TP + FN)$ and the predicted one by $(TP + FP)$.

Sensitivity and false discovery rate (FDR) [34], defined as

$$Sensitivity = \frac{TP}{TP + FN} \tag{10}$$

$$FDR = \frac{FP}{TP + FP} \tag{11}$$

can also be used to measure the detection quality, as complements to Dice coefficient.

### 5.3.2 Ablation Study

The proposed system, shown in Fig. 6, has 2 important characters for the prediction of the tumor pixel distribution and the tumor detection.

- Modulation of the asymmetry maps and the 2D histograms. The modulation functions are made adaptive to each patient case as they are generated from the original data of the case.

- Progressive prediction and cropping. The 3D data sample is cropped following each of the 3 coarse predictions of tumor pixel distribution. The final prediction is done after most of the tumor-free pixels are removed by the cropping operations.

The ablation study has been done by 2 trials related to the 2 characters. The 2 trials have been conducted with the data of all the 1251 patient cases of BraTS2021. The results comprise the similarity, measured in SSIM, between predicted and true 2D histograms and Dice score of tumor detection.

The first trial is to test if the adaptability of the modulation functions $f_{m1}$ and $f_{m2}$ contribute positively to the performance of the proposed system. To this end, the system is modified by replacing $f_{m1}$ and $f_{m2}$ by a unit-step function $u(i-0.3)$, suppressing all the elements in the lower 30% gray level section, regardless the original pixel distributions. The test result of this modified system is presented in the first part of Table 1, in comparison with that of the proposed one. One can see a significant difference between the two, proving the effectiveness of the proposed adaptive modulation functions and the necessity to apply them to achieve a good processing quality in the prediction and the detection.

In the proposed system, the prediction is done step-by-step while the regions of non-interest, i.e., tumor-free regions, are cropped out progressively, improving the density of object information and prediction quality. The 2$^{nd}$ trial has been conducted to prove the contribution of the cropping operations to the system performance. In this trial, the system is modified 3 times, resulting in 3 modified versions, shown in Fig. 25 (b), (c) and (d). In the first one, there is no cropping at all, in the second one only tumor-free axial slices are cropped out before the final prediction, and in the 3rd one there are only 2 cropping operations, instead of 3 in the proposed system. The 3 versions have been tested and the results are presented in the second half of Table 1. One can see that, the more the tumor-free regions are cropped out, the higher information density in the data to be processed in the next step, and the better the quality for the prediction and the detection. The best result is evidently given by the proposed system with 3 cropping operations.

Table 1  Similarity (SSIM) of histogram prediction and Dice coefficient of tumor detection obtained in the ablation study of Trial 1 and 2

| Trials | Descriptions | | Histogram similarity, SSIM* | | | Tumor detection, Dice* |
|---|---|---|---|---|---|---|
| | | | Axial | Coronal | Sagittal | |
| Trial 1. Modulation of the 2D histograms | Step function $u(i-0.3)$ replacing the adaptive modulations | 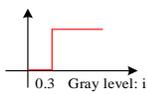 | 0.777 | 0.775 | 0.785 | 0.762 |
| | **Proposed** modulation functions $f_{m1}(i)$ and $f_{m2}(i)$ adapting to individual cases | 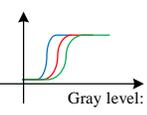 | 0.841 | 0.837 | 0.851 | 0.802 |
| Trial 2. Prediction with or without cropping operations | No cropping, Fig. 25 (b) | | 0.669 | 0.669 | 0.689 | 0.482 |
| | Only first cropping, Fig. 25 (c) | | 0.789 | 0.778 | 0.794 | 0.613 |
| | First and second cropping, Fig. 25 (d) | | 0.842 | 0.836 | 0.839 | 0.735 |
| | **Proposed** (cropping in all the 3 dimensions), Fig. 25 (a) | | 0.841 | 0.837 | 0.851 | 0.802 |

* Mean values obtained by testing the 1251 patient cases of BraTS2021

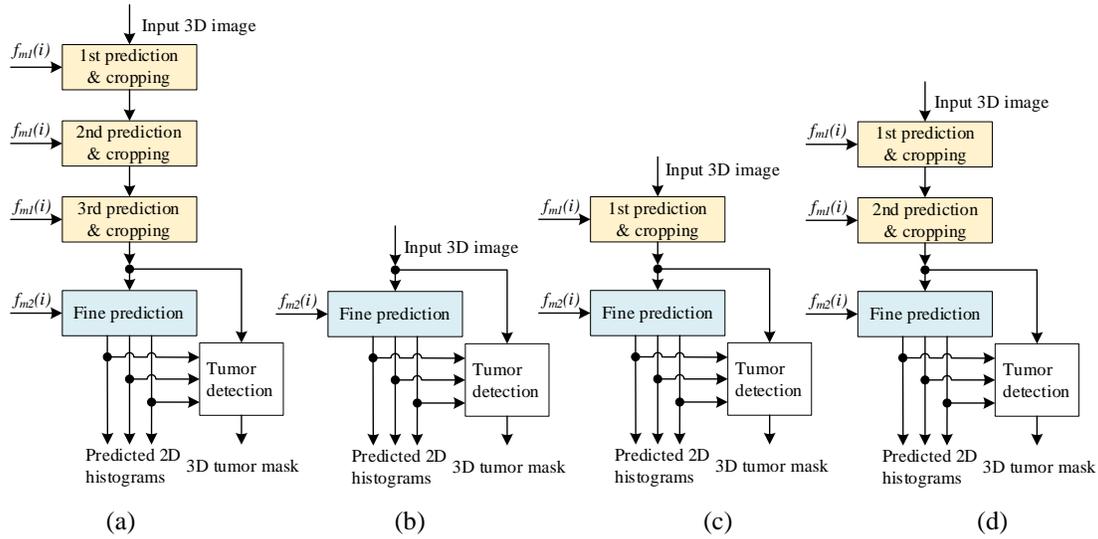

Fig. 25 Several variations of the ablation study in Trial 2.
    (a) Simplified version of the block diagram of the proposed system shown in Fig. 6.
    (b)(c)(d) Block diagrams of the ablated versions

In summary, the results of the two trials in this ablation study confirm the essentiality and the effectiveness of the proposed modulation method and the process of progressively predicting/cropping. These 2 combined make it possible to obtain a high-quality prediction and tumor detection by an extremely simple computation process.

### 5.3.3 Prediction Results

To evaluate the prediction quality of the proposed system, the similarity between the predicted 2D histogram of the tumor pixels in each of the axial, coronal and sagittal series and its ground truth has been measured on the 1251 patient cases from BraTS 2021 dataset. Large varieties of tumors appear in these 1251 cases, and some are more difficult to detect than others. The test results, presented as statistic values of CC, MSE and SSIM measures, are shown in Table 2. The following two points are observed.

- Overall, the proposed system is able to deliver predicted 2D histograms of good quality, confirmed by the overall average SSIM of 84.3% and MSE of 0.004 on the 1251 cases.

- The median SSIM value of 2D histograms, obtained from axial, coronal or sagittal series, is visibly higher than the mean value, and even the 25-quantile is around 80%. It is confirmed that the proposed prediction method yields a good result for a large majority of patient cases.

Table 2  Similarity between the predicted and true gray level distributions. The data are generated by testing the 1251 patient cases of BraTS2021 dataset

|  |  | Predicted 2D gray level distributions of tumor pixels | | | Predicted 1D gray level distributions of tumor pixels | |
|---|---|---|---|---|---|---|
|  |  | Axial | Coronal | Sagittal | 1D histogram | CDF |
| SSIM | **mean** | **0.841** | **0.837** | **0.851** | **0.773** | **0.944** |
|  | median | 0.944 | 0.942 | 0.947 | 0.911 | 0.967 |
|  | 25quantile | 0.798 | 0.796 | 0.818 | 0.728 | 0.945 |
|  | 75quantile | 0.979 | 0.978 | 0.980 | 0.954 | 0.973 |
| Correlation | **mean** | **0.878** | **0.872** | **0.887** | **0.788** | **0.958** |
|  | median | 0.956 | 0.954 | 0.958 | 0.925 | 0.971 |
|  | 25quantile | 0.856 | 0.855 | 0.874 | 0.768 | 0.960 |
|  | 75quantile | 0.982 | 0.981 | 0.983 | 0.958 | 0.974 |
| MSE | **mean** | **0.0049** | **0.0038** | **0.0028** | **0.0478** | **0.0113** |
|  | median | 0.0016 | 0.0013 | 0.0009 | 0.0157 | 0.0027 |
|  | 25quantile | 0.0006 | 0.0005 | 0.0004 | 0.0048 | 0.0007 |
|  | 75quantile | 0.0054 | 0.0041 | 0.0031 | 0.0566 | 0.0101 |

The prediction results have been applied to detect brain tumor by means of a block of very simple thresholding/filtering process. The test results of the tumor detection are presented in the following subsubsection.

### 5.3.4 Tumor Detection Results

As the proposed system does not need training, the data samples in both validation and training sets have been used to assess its quality of the brain tumor detection delivered by the proposed system. It is done in the following 2 approaches.

- Testing on the 4 validation sets from BraTS 2018, 2019, 2020 and 2021 datasets, respectively, and using the online validation tools, namely CBICA IPP [32] and Synapse [24], to get the results.

- Testing on the 4 training sets from the same datasets and measuring the detection quality with the available ground truth data. The advantage of this approach is that the test is done on a large number of data samples. For example, one can test on the 1251 patient cases in the training set, instead of 219 in the validation set, of BraTS2021 dataset. Thus, the test has been done quite comprehensively.

In total, there are 8 tests, each on a different set of patient cases. The test results, measured in Dice coefficient, Sensitivity and FDR, are summarized in Table 3, presented in 8 columns. One can see that, in all the columns, the mean Dice coefficients are higher than 0.80 and the median values are higher than 0.86. In the right-most column, given by the test performed extensively on the 1251 patient cases, the Dice coefficients of 25 quantile is 0.767, indicating that the Dice coefficients of the upper 75% of the cases are 0.767 or more. It has been confirmed that the proposed detection method, applying the results of the predicted tumor pixel distributions, is very effective to detect most brain tumors, despite the vast variations in their locations, shapes, sizes and texture patterns in the tumor areas.

It should be underlined that, as the parameters of the proposed system are not determined by training, neither the problem of randomness in training nor the problem of reproducibility could arise. Hence the performance is robust and reliable.

Table 3  Dice coefficient, sensitivity and FDR of tumor detected by the proposed system

| Flair mono-modality input | | Test on validation set (assessed by online portal) | | | | Test on training set | | | |
|---|---|---|---|---|---|---|---|---|---|
| | | BraTS 2018 | BraTS 2019 | BraTS 2020 | BraTS 2021 | BraTS 2018 | BraTS 2019 | BraTS 2020 | BraTS 2021 |
| Number of patient cases | | 66 | 125 | 125 | 219 | 285 | 335 | 369 | **1251** |
| Dice | mean | **0.843** | **0.816** | **0.816** | **0.812** | **0.814** | **0.816** | **0.818** | **0.802** |
| | median | 0.884 | 0.881 | 0.881 | 0.874 | 0.869 | 0.873 | 0.872 | 0.876 |
| | 25quantile | 0.807 | 0.771 | 0.771 | 0.769 | 0.772 | 0.786 | 0.786 | 0.767 |
| | 75quantile | 0.914 | 0.917 | 0.917 | 0.915 | 0.913 | 0.914 | 0.914 | 0.920 |
| Sensitivity | mean | **0.850** | **0.819** | **0.819** | **0.824** | **0.825** | **0.835** | **0.834** | **0.827** |
| | median | 0.896 | 0.877 | 0.877 | 0.885 | 0.881 | 0.896 | 0.893 | 0.904 |
| | 25quantile | 0.815 | 0.756 | 0.756 | 0.777 | 0.777 | 0.787 | 0.778 | 0.785 |
| | 75quantile | 0.940 | 0.928 | 0.928 | 0.941 | 0.944 | 0.949 | 0.948 | 0.955 |
| FDR | mean | **0.127** | **0.138** | **0.138** | **0.154** | **0.150** | **0.154** | **0.150** | **0.162** |
| | median | 0.102 | 0.097 | 0.097 | 0.102 | 0.091 | 0.100 | 0.092 | 0.101 |
| | 25quantile | 0.175 | 0.176 | 0.176 | 0.188 | 0.190 | 0.190 | 0.183 | 0.190 |
| | 75quantile | 0.044 | 0.033 | 0.033 | 0.048 | 0.037 | 0.042 | 0.042 | 0.049 |

## 5.4 Performance Comparison

The performance of the proposed system, in terms of detection quality, has been compared with that of other systems reported in reputed journals. Two issues should be noted in this performance comparison.

- To make the comparison meaningful, the tests results should be produced on the same data samples. Hence it is important to specify the testing data samples in every comparison.

- It is evident that the proposed system requires much less computation with respect to the other systems, CNN or non-CNN, that have been reported in literature in recent years.

The comparison results of the proposed system with 2 non-CNN systems are presented in Table 4. These 2 systems were chosen for the comparison because their detection quality was assessed with BraTS dataset and the test conditions were similar to those of the proposed system.

Table 5 summarizes the comparison of the detection quality between the proposed system and a number of CNNs reported recently. As their tests were on mono-modality data samples from BraTS 2018 and 2019 datasets, the results are comparable to those of the proposed system.

The data presented in Table 4 and Table 5 demonstrate that the brain tumor detection performed by the proposed system is of good quality. Its dice scores are comparable to those given by CNN systems, if not better. It should, however, be noted that the tests of the proposed system have been done on a much larger number of patient cases with more varieties, and moreover the results are completely reproducible.

It should also be noted that the computation complexity of the proposed system is extremely low. The detection can be performed in an ordinary laptop or desktop. It takes only 0.85 seconds to process a patient case on a laptop of i7-11800H CPU with clock of 4.6 GHz.

Table 4  Comparison of the results of the proposed system with those of other systems without CNN

| Systems | Dataset | # patient cases testing | Input modalities | Dice | Sensitivity |
|---|---|---|---|---|---|
| Lim et al. 2018 [14] | BraTS2013 | 20$^{(H)}$ | T1c, T2 | 0.701 | 0.856 |
| Lim et al. 2018 [14] | BraTS2013 | 10$^{(L)}$ | T1c, T2 | 0.692 | 0.717 |
| Bonte et al. 2018 [17] | BraTS2017 | 210$^{(H)}$ | Flair, T1c | 0.762 | N.A. |
| Bonte et al. 2018 [17] | BraTS2017 | 75$^{(L)}$ | Flair, T1c | 0.656 | N.A. |
| **Proposed** | BraTS2013 | 20$^{(H)}$ | Flair | **0.777** | **0.759** |
| **Proposed** | BraTS2013 | 10$^{(L)}$ | Flair | **0.709** | **0.866** |
| **Proposed** | BraTS2017 | 210$^{(H)}$ | Flair | **0.829** | **0.844** |
| **Proposed** | BraTS2017 | 75$^{(L)}$ | Flair | **0.772** | **0.772** |
| **Proposed** | BraTS2021 | **1251** | Flair | **0.802** | **0.827** |

(H): High-grade glioma
(L): Low-grade glioma

Table 5  Comparison of the results of the proposed system with those of CNN systems

| Systems | Dataset | # patient cases testing | Input modality | Dice | Sensitivity |
|---|---|---|---|---|---|
| Wu et al. 2021 [3] | BraTS2018 | 143* | T2 | 0.619 | 0.613 |
| Zhou et al. 2021 [4] | BraTS2018 | 57* | Flair | 0.737 | N.A. |
| Yang et al. 2022 [5] | BraTS2018 | 95* | Flair | 0.842 | N.A. |
| Zhou et al. 2021 [4] | BraTS2019 | 67* | Flair | 0.743 | N.A. |
| **Proposed** | BraTS2018 | **66**** | Flair | **0.843** | **0.850** |
| **Proposed** | BraTS2019 | **125**** | Flair | **0.816** | **0.819** |
| **Proposed** | BraTS2021 | **1251** | Flair | **0.802** | **0.827** |

\* Test on cases taken randomly from the training set
\*\* Test on the validation set, results assessed by CBICA Image Processing Portal [32]

## 6. Conclusion

The challenges in brain tumor detection, like in all kinds of object detections, are often related to the extremely low density of object information in the input data and the enormous variations of the objects. In this paper, we have proposed a system that predicts the gray level distributions of tumor pixels, i.e., pixels in the tumor regions, of a 3D MRI brain scan of Flair modality, and detects precisely tumor locations in the 3D scan. In this work, we have proposed (i) 2D histogram presentations of the data in the axial, coronal and sagittal slice series of a 3D image, comprehending the distributions of the gray levels of the pixels with their locations, (ii) extraction of brain tumor information by exploiting the left-right asymmetry of a brain structure, (iii) histogram modulation, on a case-by-case basis, to enhance the structural asymmetry related to the presence of the tumor and to attenuate that due to non-pathological causes, (iv) step-by-step prediction of tumor pixel distribution, accompanied by step-by-step cropping out areas of non-interest to improve the signal density, and (v) tumor detection process consisting of a simple thresholding, based on the prediction results, and a low-pass filtering for morphological purpose.

The proposed system does not need training. It has been tested extensively with the data of more than one thousand patient cases in BraTS 2018~21 datasets. The test results demonstrate that, with the input data of only Flair modality, the predicted 2D histograms have a high degree of similarity with respect to the true ones. Also, the tumor detection performed by the system is also of high-quality. It is worth mentioning that the good performance of the proposed system has been achieved at an extremely low computation cost that may be negligible with respect to those of other state-of-the-art systems.

## Acknowledgements

This work was supported in part by Digital Research Alliance of Canada and in part by the Natural Sciences and Engineering Research Council (NSERC) of Canada.